\documentclass[11pt]{article}
\usepackage{a4p,epsfig}
\newcommand{\Zz}{\ensuremath{{\mathrm{Z}^0}}}
\newcommand{\alphas}{\ensuremath{\alpha_\mathrm{s}}}
\newcommand{\bq}{\ensuremath{\mathrm{b}}}
\newcommand{\cq}{\ensuremath{\mathrm{c}}}
\newcommand{\cc}{\ensuremath{\mathrm{c\overline{c}}}}
\newcommand{\bb}{\ensuremath{\mathrm{b\overline{b}}}}
\newcommand{\qq}{\ensuremath{\mathrm{q\overline{q}}}}
\newcommand{\qqg}{\ensuremath{\mathrm{q\overline{q}g}}}
\newcommand{\qqgg}{\ensuremath{\mathrm{q\overline{q}gg}}}
\newcommand{\qqbb}{\ensuremath{\mathrm{q\overline{q}b\overline{b}}}}
\newcommand{\bbbb}{\ensuremath{\mathrm{b\overline{b}b\overline{b}}}}
\newcommand{\qqcc}{\ensuremath{\mathrm{q\overline{q}c\overline{c}}}}
\newcommand{\epem}{\ensuremath{\mathrm{e}^{+}\mathrm{e}^{-}}}
\newcommand{\gcc}{\ensuremath{g_{\cc}}}
\newcommand{\gbb}{\ensuremath{g_{\bb}}}
\newcommand{\gfb}{\ensuremath{g_{\mathrm{4b}}}}
\newcommand{\gtocc}{\ensuremath{\mathrm{g}\to{\cc}}}
\newcommand{\gtobb}{\ensuremath{\mathrm{g}\to{\bb}}}

\newcommand{\Rc}{\ensuremath{R_{\mathrm{c}}}}
\newcommand{\Rb}{\ensuremath{R_{\mathrm{b}}}}

\newcommand{\Rbfour}{\ensuremath{R_{\mathrm{b}}^{\mathrm{4-jet}}}}
\newcommand{\yc}[1]{\ensuremath{y_{#1}}}
\newcommand{\ycut}{\ensuremath{y_{\mathrm{cut}}}}
\newcommand{\Jetset}{\mbox{{\sc Jetset~7.4}}}
\newcommand{\Newpythia}{\mbox{{\sc Pythia~6.130}}}
\newcommand{\Ariadne}{\mbox{{\sc Ariadne~4.08}}}
\newcommand{\Herwig}{\mbox{{\sc Herwig~5.91}}}
\newcommand{\WPHACTnv}{\mbox{{\sc Wphact}}}
\newcommand{\WPHACT}{\mbox{{\sc Wphact~1.3}}}
\newcommand{\kT}{\ensuremath{k_{\perp}}}
\newcommand{\ezero}{\mbox{{\sc JADE E0}}}
\newcommand{\vnn}[1]{\ensuremath{{N\!N}_{#1}}}
\newcommand{\bz}{\ensuremath{\alpha_{12-34}}}
\newcommand{\ebb}{\ensuremath{\epsilon_{\qqbb}}}
\newcommand{\efb}{\ensuremath{\epsilon_{\bbbb}}}
\newcommand{\ecc}{\ensuremath{\epsilon_{\qqcc}}}
\newcommand{\eqq}{\ensuremath{\epsilon_{\qq}}}
\newcommand{\dgbb}{\ensuremath{\Delta g_{\bb}}}
\newcommand{\dgfb}{\ensuremath{\Delta g_{\mathrm{4b}}}}
\newcommand{\geve}{\ensuremath{\mathrm{GeV}}}
\newcommand{\gevp}{\ensuremath{\mathrm{GeV}/c}}
\newcommand{\gevm}{\ensuremath{\mathrm{GeV}/c^2}}
\newcommand{\rbval}{\ensuremath{0.21664}}

\newcommand{\rcval}{\ensuremath{0.1724}}
\newcommand{\gccval}{\ensuremath{3.20}} 
\newcommand{\dgccval}{\ensuremath{0.43}} 

\newcommand{\nbins}{\ensuremath{26}}
\newcommand{\gbbjs}{\ensuremath{2.5}} 
\newcommand{\gfbjs}{\ensuremath{0.4}} 
\newcommand{\ngbb}{\ensuremath{3.07}}       
\newcommand{\dngbb}{\ensuremath{0.53}}      
\newcommand{\ngfb}{\ensuremath{0.36}}       
\newcommand{\dngfb}{\ensuremath{0.17}}      
\newcommand{\gfgbcorr}{\ensuremath{+0.007}} 
\newcommand{\fracfb}{\ensuremath{0.116}}    
\newcommand{\dfracfb}{\ensuremath{0.060}}   
\newcommand{\ngbbjade}{\ensuremath{3.11}}     
\newcommand{\dngbbjade}{\ensuremath{0.59}}    
\newcommand{\ngfbjade}{\ensuremath{0.45}}     
\newcommand{\dngfbjade}{\ensuremath{0.17}}    
\newcommand{\ncanjade}{\ensuremath{207}} %
\newcommand{\dmcstat}{\ensuremath{0.28}}   
\newcommand{\dfmcstat}{\ensuremath{0.09}}  
\newcommand{\cmcstat}{\ensuremath{-0.07}}  
\newcommand{\dnorm}{\ensuremath{0.18}}    
\newcommand{\dfnorm}{\ensuremath{0.03}}   
\newcommand{\cnorm}{\ensuremath{+1}}
\newcommand{\dtracking}{\ensuremath{0.23}}   
\newcommand{\dftracking}{\ensuremath{0.15}}  
\newcommand{\ctracking}{\ensuremath{+1}} 
\newcommand{\dbmass}{\ensuremath{0.23}}  
\newcommand{\dfbmass}{\ensuremath{0.03}} 
\newcommand{\cbmass}{\ensuremath{+1}}
\newcommand{\ngbbjs}{\ensuremath{2.35}}  
\newcommand{\dngbbjs}{\ensuremath{0.41}} 
\newcommand{\ngfbjs}{\ensuremath{0.34}}  
\newcommand{\dngfbjs}{\ensuremath{0.17}} 
\newcommand{\ngbbhw}{\ensuremath{3.38}}  
\newcommand{\dngbbhw}{\ensuremath{0.59}} 
\newcommand{\ngfbhw}{\ensuremath{0.40}}  
\newcommand{\dngfbhw}{\ensuremath{0.19}} 
\newcommand{\ngbbwp}{\ensuremath{3.22}}  
\newcommand{\dngbbwp}{\ensuremath{0.58}} 
\newcommand{\ngfbwp}{\ensuremath{0.31}}  
\newcommand{\dngfbwp}{\ensuremath{0.15}} 
\newcommand{\ngbbar}{\ensuremath{3.62}}  
\newcommand{\dngbbar}{\ensuremath{0.64}} 
\newcommand{\ngfbar}{\ensuremath{0.36}}  
\newcommand{\dngfbar}{\ensuremath{0.19}} 
\newcommand{\dmodel}{\ensuremath{0.72}}  
\newcommand{\dfmodel}{\ensuremath{0.17}} 
\newcommand{\cmodel}{\ensuremath{+1}}
\newcommand{\dgbbgcc}{\ensuremath{0.05}}  
\newcommand{\dfgbbgcc}{\ensuremath{0.01}} 
\newcommand{\cgbbgcc}{\ensuremath{+1}}
\newcommand{\delrgbbrgcc}{\ensuremath{-0.125}} 
\newcommand{\delrgfbrgcc}{\ensuremath{-0.259}} 
\newcommand{\dgbbrb}{\ensuremath{0.07}}  
\newcommand{\dfgbbrb}{\ensuremath{0.02}} 
\newcommand{\cgbbrb}{\ensuremath{+1}}
\newcommand{\devrb}{\ensuremath{2.4\%}} 
\newcommand{\dcprod}{\ensuremath{0.02}}   
\newcommand{\dfcprod}{\ensuremath{0.01}}  
\newcommand{\ccprod}{\ensuremath{-0.97}}  
\newcommand{\dbprod}{\ensuremath{0.11}}   
\newcommand{\dfbprod}{\ensuremath{0.01}}  
\newcommand{\cbprod}{\ensuremath{+0.70}}  
\newcommand{\dcfrag}{\ensuremath{0.07}}   
\newcommand{\dfcfrag}{\ensuremath{0.03}}  
\newcommand{\ccfrag}{\ensuremath{-0.52}} 
\newcommand{\dbfrag}{\ensuremath{0.39}}   
\newcommand{\dfbfrag}{\ensuremath{0.10}}  
\newcommand{\cbfrag}{\ensuremath{+0.86}} 
\newcommand{\dctau}{\ensuremath{0.05}}  
\newcommand{\dfctau}{\ensuremath{0.01}} 
\newcommand{\cctau}{\ensuremath{+0.80}} 
\newcommand{\dbtau}{\ensuremath{0.11}}  
\newcommand{\dfbtau}{\ensuremath{0.05}} 
\newcommand{\cbtau}{\ensuremath{+0.09}} 
\newcommand{\dcmult}{\ensuremath{0.03}}  
\newcommand{\dfcmult}{\ensuremath{0.01}} 
\newcommand{\ccmult}{\ensuremath{+0.96}} 
\newcommand{\dbmult}{\ensuremath{0.11}}  
\newcommand{\dfbmult}{\ensuremath{0.03}} 
\newcommand{\cbmult}{\ensuremath{+1}}
\newcommand{\dngbbsys}{\ensuremath{0.97}}     
\newcommand{\dngfbsys}{\ensuremath{0.27}}     
\newcommand{\gfgbcorrsys}{\ensuremath{+0.78}} 
\newcommand{\dfracfbsys}{\ensuremath{0.065}}  
\newcommand{\ndatatot}{\ensuremath{3.35}~million}
\newcommand{\nmctot}{\ensuremath{14}~million}

\newcommand{\ngbbtotpy}{\ensuremath{150\,000}}
\newcommand{\ncctot}{\ensuremath{1.9}~million}
\newcommand{\nbbtot}{\ensuremath{4.3}~million}
\newcommand{\nfourjet}{\ensuremath{443\,334}}
\newcommand{\ffourjet}{\ensuremath{13.0\%}}
\newcommand{\ffourjetmcmh}{\ensuremath{11.5\%}}
\newcommand{\ntwotwo}{\ensuremath{235\,201}}
\newcommand{\ntwotwomc}{\ensuremath{240\,774}}
\newcommand{\nthreeone}{\ensuremath{208\,133}}
\newcommand{\nthreeonemc}{\ensuremath{202\,560}}
\newcommand{\ncand}{\ensuremath{221}}
\newcommand{\nexpect}{\ensuremath{217}}
\newcommand{\dnexpect}{\ensuremath{7}}
\newcommand{\effbb}{\ensuremath{0.64}}  
\newcommand{\deffbb}{\ensuremath{0.02}} 
\newcommand{\efffb}{\ensuremath{1.64}}  
\newcommand{\defffb}{\ensuremath{0.07}} 
\newcommand{\fourjetpur}{\ensuremath{1.2\%}}
\newcommand{\fourjetdeficit}{\ensuremath{11.7}}   
\newcommand{\dfourjetdeficit}{\ensuremath{0.2}}   
\newcommand{\dfourjetdeficitcc}{\ensuremath{0.8}} 
\newlength{\picwi}
\begin{document}
\begin{titlepage}
\begin{center}
{\large EUROPEAN ORGANIZATION FOR NUCLEAR RESEARCH}
\end{center}\bigskip

\begin{flushright}
       CERN-EP-2000-123 \\
       7th June 2000
\end{flushright}
\bigskip\bigskip\bigskip\bigskip
\begin{center}\huge\bf\begin{boldmath}
Production rates of

\bb{} quark pairs from gluons and \bbbb{} events
\\[1.5mm]

in hadronic \Zz{} decays
\end{boldmath}\end{center}
\bigskip\bigskip
\begin{center}{\LARGE The OPAL Collaboration
}\end{center}\bigskip\bigskip
\bigskip
\bigskip
\begin{abstract}
The rates are measured per hadronic \Zz{} decay for gluon splitting to
\bb{} quark pairs, \gbb{}, and of events containing two \bb{} quark
pairs, \gfb{}, using a sample of four-jet events selected from 
data collected with the OPAL detector.
Events with an enhanced signal of gluon splitting to \bb{} quarks 
are selected if two of the jets are close in phase-space
and contain detached secondary vertices. For the event sample containing two
\bb{} quark pairs, three of the four jets are required to have a
significantly detached secondary vertex. Information from the event topology
is combined in a likelihood fit to extract the
values of \gbb{} and \gfb{}, namely
\begin{eqnarray*}
 \gbb & = & (\ngbb \pm \dngbb \mathrm{(stat)} \pm 
   \dngbbsys\mathrm{(syst)} )\times 10^{-3}, \\
 \gfb & = & (\ngfb \pm \dngfb \mathrm{(stat)} \pm 
   \dngfbsys\mathrm{(syst)} )\times 10^{-3}.
\end{eqnarray*}
\end{abstract}
\bigskip\bigskip
\begin{center}\large Submitted to Eur.~Phys.~J.~{\bf C}\end{center}
\bigskip
\end{titlepage}

\begin{center}{\Large        The OPAL Collaboration
}\end{center}\bigskip
\begin{center}{
G.\thinspace Abbiendi$^{  2}$,
K.\thinspace Ackerstaff$^{  8}$,
C.\thinspace Ainsley$^{  5}$,
P.F.\thinspace Akesson$^{  3}$,
G.\thinspace Alexander$^{ 22}$,
J.\thinspace Allison$^{ 16}$,
K.J.\thinspace Anderson$^{  9}$,
S.\thinspace Arcelli$^{ 17}$,
S.\thinspace Asai$^{ 23}$,
S.F.\thinspace Ashby$^{  1}$
% ...
}\end{center}\bigskip
\bigskip
$^{  1}$School of Physics and Astronomy, University of Birmingham,
Birmingham B15 2TT, UK
\newline
$^{  2}$Dipartimento di Fisica dell' Universit\`a di Bologna and INFN,
I-40126 Bologna, Italy
\newline
$^{  3}$Physikalisches Institut, Universit\"at Bonn,
D-53115 Bonn, Germany
\newline
% ...
$^{  5}$Cavendish Laboratory, Cambridge CB3 0HE, UK
\newline
% ...
$^{  8}$CERN, European Organisation for Nuclear Research,
CH-1211 Geneva 23, Switzerland
\newline
$^{  9}$Enrico Fermi Institute and Department of Physics,
University of Chicago, Chicago IL 60637, USA
\newline
% ...
$^{ 16}$Department of Physics, Schuster Laboratory, The University,
Manchester M13 9PL, UK
\newline
$^{ 17}$Department of Physics, University of Maryland,
College Park, MD 20742, USA
\newline
% ...
$^{ 22}$Department of Physics and Astronomy, Tel Aviv University,
Tel Aviv 69978, Israel
\newline
$^{ 23}$International Centre for Elementary Particle Physics and
Department of Physics, University of Tokyo, Tokyo 113-0033, and
Kobe University, Kobe 657-8501, Japan
\newline
% ...
Budapest, Hungary.
\bigskip
\section{Introduction}

Bottom quark pairs in \Zz{} decays can be produced either directly via
$\Zz\to\bb$ or indirectly, when a gluon is radiated from a quark and then
splits into a \bb{} quark pair. A special case consists of events with
both direct and indirect \bq{} quark production, $\Zz\to\bb\gtobb\bb$.
The rates \gbb{} and \gfb{} per hadronic \Zz{} decay, for the
reactions $\Zz\to\qqg$ with $\gtobb$ and
$\Zz\to\bbbb{}$  are sensitive to both the \bq{}
quark mass and the strong coupling constant $\alpha_s$. Hence measurements
of these rates are tests of the theory of Quantum Chromodynamics
(QCD). Events with gluon splitting into \bb{} quark pairs are an
important background for the measurement of \Rb{}, the fraction of
hadronic \Zz{} decays into \bb{} quark pairs. Therefore, a more
precise determination of \gbb{} might lead to a reduction of the
systematic uncertainty in \Rb{}.

The rate \gbb{} has been calculated in \cite{seymour95,miller98}, 
including the re-summation of leading logarithmic terms.
Those authors point out that the parton-shower approach as
implemented e.g.~in \Jetset{} \cite{jetset} is a good approximation.
Numerical calculations of \gbb{} are given in
\cite{seymour95,miller98,frixione97},
predicting a rate in the range $\gbb^\mathrm{theor}=(1.8 -
2.9)\times 10^{-3}$, depending on the \bq{} quark mass and the strong
coupling constant.

An estimate of \gfb{} can be obtained from the rate for direct \bb{}
production, \Rb{}, multiplied by the rate of indirect \bb{} production,
\gbb{}. This simple picture is modified because the phase-space
for two \bb{} quark pairs is smaller than that for two light and two \bq{}
quarks. The interference between secondary \bb{} production 
and primary \bb{} production cancels to zero at leading
order in \alphas{}, except for the \bbbb{} final state
\cite{seymour95,kniehl}. In \cite{seymour95}, the contribution of this
interference term is shown to be less than $0.2\%$ of \gbb{}.

Measurements of \gbb{} using four-jet
final states have been reported by the DELPHI and ALEPH
collaborations, with the results
$\gbb=(2.1\pm 1.1(\mathrm{stat})\pm 0.9(\mathrm{syst}))\times 10^{-3}$
\cite{delphi} and 
$\gbb=(2.77\pm 0.42(\mathrm{stat})\pm 0.57(\mathrm{syst}))\times
10^{-3}$ \cite{aleph}, respectively,
in agreement with theoretical predictions.
A measurement of the \bb\bb{} final state in
a three jet topology has recently been presented by
DELPHI~\cite{delphifour}, with the result $\gfb=(0.60\pm
0.19(\mathrm{stat})\pm 0.14(\mathrm{syst}))\times 10^{-3}$. This result
is translated into a gluon splitting rate
$\gbb=(3.3\pm1.0(\mathrm{stat})\pm0.8(\mathrm{syst}))\times 10^{-3}$,
using a tree-level QCD calculation for the ratio $\gbb /\gfb$. The
calculation was carried out by DELPHI~\cite{delphifour}, using the
\WPHACT{} Monte Carlo generator \cite{accomando97}.

In this analysis, decays of the \Zz{} into four-jet final states are
investigated. Jets from \bq{} or $\overline{\bq}$ quarks are
identified by reconstructing 
secondary decay vertices.
The invariant mass of \bb{} quark pairs originating from gluons tends
to peak just above threshold.
This leads to a small relative momentum of the two
\bq{} hadrons produced in the fragmentation process.
By contrast, directly produced \bb{} quark pairs have high
invariant masses, since
they carry a large part of the \Zz{} energy. This and other
characteristics are used to select event samples enriched in the
process \gtobb. An angular correlation defined similarly to the
Bengtsson-Zerwas angle \cite{bzerwas} is used to further differentiate
between \qq\qq{} and \qqgg{} final states.
In addition, events with three reconstructed secondary vertices are
selected. They are used to measure the \gfb{} rate.

\section{Event selection and reconstruction}

\subsection{The OPAL detector}

The OPAL detector is described in detail elsewhere \cite{opaldet}.
Only a brief description of the detector elements relevant to
this analysis is given here. Charged tracks are reconstructed in the central
tracking system. It consists of a silicon microvertex detector, a vertex
drift chamber equipped with axial and stereo wires, a large jet chamber and
$z$ chambers\footnote{The OPAL coordinate system is
defined with positive $z$-axis along the electron beam direction, the $x$ axis
pointing to the center of the LEP accelerator ring and the $y$-axis
normal to the $x$--$z$ plane. The polar and
azimuthal angles are denoted by  $\theta$ and $\phi$, respectively.}.
A solenoid providing a uniform magnetic field of $0.435$
T parallel to the $z$-axis surrounds the central tracking system.
The silicon microvertex detector \cite{opalsi} has two layers which
measure tracks in $(r,\phi)$.  This detector was
upgraded in 1993 to provide a precise measurement of the $z$
coordinate \cite{opalsi1}.  Before this detector was upgraded again in
1995 for the high energy operation in the LEP~2 programme
\cite{opalsi2} the inner layer 
covered the range $\vert\cos(\theta)\vert<0.83$ and the outer layer
the range $\vert\cos(\theta)\vert<0.77$.
The vertex chamber extends over the range $\vert\cos\theta\vert<0.95$.
The coil is surrounded by scintillators for time-of-flight
measurements and a barrel lead-glass electromagnetic calorimeter.
Including the endcap electromagnetic calorimeter, the lead-glass
blocks cover the range $\vert\cos\theta\vert<0.98$.
The magnetic return yoke is instrumented with streamer
tubes and serves as a hadron calorimeter. The return yoke in turn
is surrounded by muon chambers.

\subsection{Event selection and reconstruction} 

The analysis uses data taken with the OPAL detector in the
years 1992--1995 on or near the \Zz{} resonance.
Hadronic \Zz{} decays are selected with an
efficiency of $98.4\%$, as described in \cite{rmhsel}.
Only events with the tracking system and the electromagnetic calorimeter
fully operational are used in this analysis. A total number of
\ndatatot{} hadronic events are selected.
In these events well-measured charged tracks are used with a momentum
$p_{\mathrm{t}}>0.15\,\gevp$ in the $(x,y)$ plane, and clusters in the
electromagnetic calorimeter with energies above $0.1\,\geve$
($0.25\,\geve$) in the barrel (end-cap) region.
The energies of clusters pointed to by charged tracks are corrected for
double counting by subtracting the energy deposition expected from
the track momentum \cite{mt}.

\subsection{Simulated events}

A total number of \ngbbtotpy{} events generated with \Newpythia{}
\cite{pythia99} and \nmctot{} events generated with
\Jetset{}~\cite{jetset} are used to evaluate the efficiencies for
signal and background, respectively. Within the \Jetset{} sample,
\nbbtot{} (\ncctot{}) events  were generated in special runs with a
primary \bb{} (\cc{}) quark pair to increase the statistical
significance of the description of background processes with secondary
decay vertices. To study the \gtobb{} signal process \ngbbtotpy{}
events with gluon splitting to \bb{} in the parton shower were
generated with \Newpythia{}. This generator is used with a special
option for gluon splitting to massive quarks, as explained in Section
\ref{textmodel}.  All signal and background events were passed through a
complete simulation of the OPAL detector \cite{gopal}.

The simulated events are weighted to
correspond to the measured values of \cc{} and \bb{}
production,  $\Rc=\rcval$ and $\Rb =\rbval$, given in \cite{lepew98}.
The rate of gluon splitting to \cc{} pairs is set to
$\gcc =\gccval\times 10^{-2}$ as measured by OPAL \cite{gtocc}.
The ratio of the number of events with primary produced \bq{} quarks
and \gtocc{} to the total number of events with \gtocc{} is fixed to
the value predicted by \Jetset{}.

In the following, simulated events with gluon splitting to \bb{}
signal are referred to as \gtobb{}.
They are further differentiated into events with a primary charm or
light quark, referred to as \qqbb{} $(\mathrm{q}=\mathrm{udsc})$, and
events with two \bb{} quark pairs, referred to as \bbbb{}.

Background to this analysis consists of various sources of four jet
events. The dominant four-jet process is the production of two gluons
in addition to the primary quarks, either from double Bremsstrahlung or
from the triple gluon vertex. About $7\%$ of the four-jet events are
expected to be from gluon-splitting to a quark antiquark pair
\cite{delphicol}.

As this analysis is based on the identification of secondary decay vertices,
there are two sources of background which have to be studied more
carefully: events with gluon splitting to \cc{} accompanied by any
flavor of primary quark, and four-jet events with a primary \bb{} 
quark pair but without gluon splitting to heavy quarks. They are
referred to as $\bb\mathrm{xx}$ $(\mathrm{x}=\mathrm{guds})$.

For comparisons to the data, the rate of gluon splitting to \bb{}
pairs is set to $\gbb = \gbbjs \times 10^{-3}$ and the rate of \bbbb{}
events is set to $\gfb = \gfbjs \times 10^{-3}$. Note that
the rate \gbb{} includes \bbbb{} events. The fit procedure, described
in Section~\ref{textmaxlfit}, is completely independent of these two numbers.

\subsection{The four-jet selection}
\label{textfourjet}

\begin{figure}[t]
\begin{center}
\epsfig{file=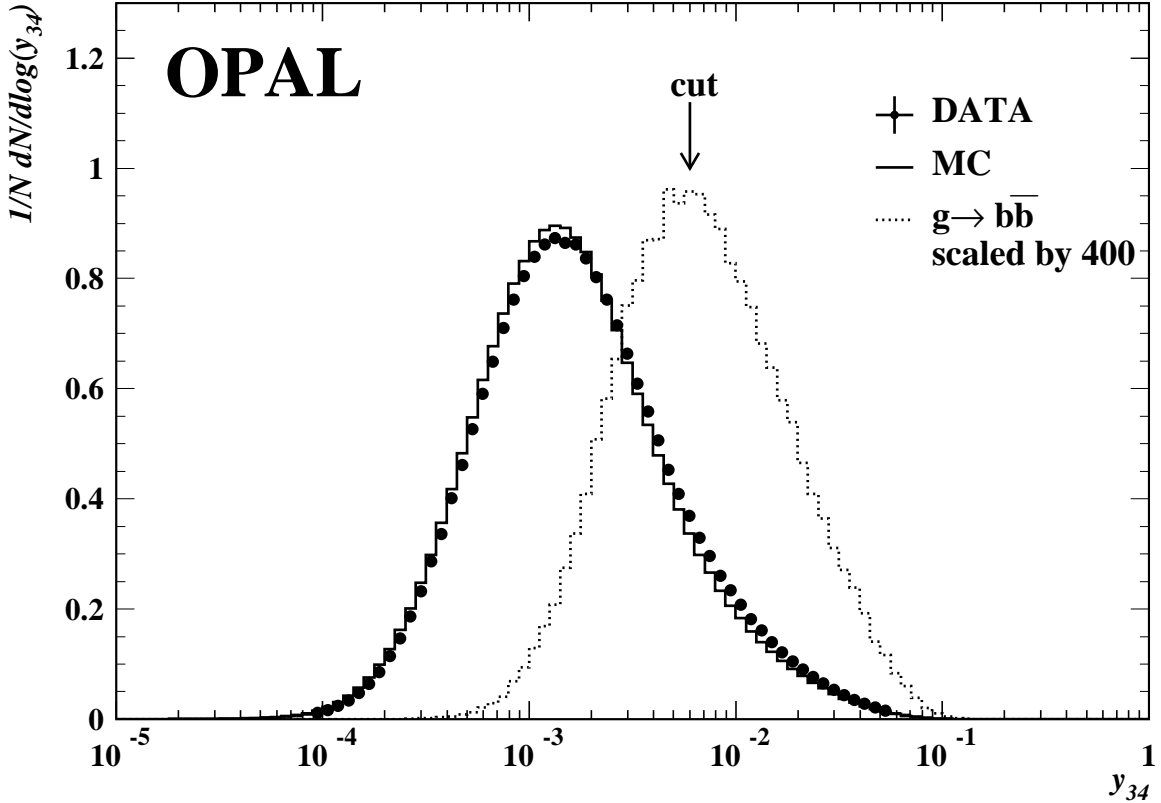,width=\picwi}
\end{center}
\caption{\it\label{figy34}
Normalized distributions of the jet resolution
parameter \yc{34}. The data are shown with dots, the
shape predicted from Monte Carlo simulations with a solid line and
the shape predicted for the signal events as a dotted line.
}
\end{figure}

The four-momenta of the selected tracks and clusters are combined to
form four jets, using the \kT{} (Durham) algorithm \cite{durham}.
The value \ycut{} at which an event makes a transition
between a three-jet and a four-jet assignment is denoted \yc{34}.
Figure \ref{figy34} shows the normalized distribution of the
quantity \yc{34} for events that are selected as hadronic \Zz{} decays,
thereby comparing the data to the Monte Carlo prediction. In addition, the
distribution predicted for \gtobb{} events is shown, scaled by a factor of 400.
These events populate mainly the region of high \yc{34}.
A cut $\yc{34}>0.006$ is made to define the four-jet sample. This cut
rejects nearly $90\%$ of the background events, while retaining about $60\%$
of the signal events. The estimated signal fraction assuming a signal rate
$\gbb=\gbbjs\times 10^{-3}$ is $\fourjetpur$.
In the data $\nfourjet$ events are selected, corresponding to
$\ffourjet$ of all hadronic \Zz{} decays.
In the Monte Carlo simulation only $\ffourjetmcmh$ of the events are
selected, because the prediction is slightly shifted to lower values
of \yc{34} with respect to the data. This corresponds to a deficit of
$(\fourjetdeficit\pm\dfourjetdeficit\pm\dfourjetdeficitcc)\%$
four-jet events in the simulation compared to the data,
where the first uncertainty is due
to the finite number of events in the data and the simulation, the
second is from the uncertainty in the rate of gluon splitting to \cc{},
$\gcc=(\gccval\pm\dgccval)\times 10^{-2}$ \cite{gtocc}.
This deficit is attributed to missing higher orders in the parton
shower simulation.
To deal with this normalization problem, the number of simulated events is
normalized to the number of four-jet events throughout this
analysis, instead of normalizing to the number of hadronic \Zz{}
decays. The stability of this analysis with
respect to variations of the cut in \yc{34}, affecting this
normalization, is discussed in Section~\ref{textxcheck}.

\subsection{Vertex reconstruction}

A primary vertex is reconstructed for each event as described
in \cite{opalrb97}. Then for each jet
a secondary vertex is reconstructed, using the full 
three-dimensional tracking information \cite{opalrb98}.
All tracks in the jet satisfying additional quality cuts on the momentum
$p>0.5\,\gevp$, the distance to the fitted primary vertex
in the $r/\phi$ plane $d_0<0.3\,\mathrm{cm}$ and the uncertainty on $d_0$ of
$\sigma_{d_0}<0.1\,\mathrm{cm}$ are fitted to a common vertex. The
relatively high momentum cut results in an increased contribution of
tracks from the decay of \bq{} hadrons.
Tracks with a large $\chi^2$ contribution are removed from the vertex
and the fit is repeated until the $\chi^2$ 
contribution of each track is smaller than 4. 
If three or more tracks remain, the secondary vertex is accepted.

The decay length  $l$ and its error $\sigma$ are calculated as
the distance of the primary to the secondary vertex. The direction is
constrained to be parallel to the jet axis. The decay length $l$ is positive 
if the angle between the jet axis and the
vector pointing from the primary to the secondary vertex is less than
90 degrees, negative otherwise. Vertices with $l>0$ are used to
identify \bq{} hadrons. In this analysis, the following variables are used
to identify secondary vertices originating from the decay of \bq{} hadrons:
\begin{itemize}
\item The decay length significance $l/\sigma$, calculated from the
decay length $l$ and the experimental uncertainty $\sigma$ on $l$.
Secondary vertices significantly separated from the primary vertex are
selected with a cut $l/\sigma>3$.
\item For the vertices surviving the $l/\sigma>3$ cut the output \vnn{}
of a neural network is calculated. This neural network has been
trained to separate vertices originating from b hadron decays from
those in charm or light quark events.
\end{itemize}
The neural network was developed for the OPAL \Rb{}
analysis \cite{opalrb98}.
It has five inputs:
the decay length significance $l/\sigma$,
the decay length $l$, the number of tracks in the secondary vertex
$n_s$, the reduced decay length $l_r/\sigma_r$, where one well-defined
track \cite{opalrb98} has
been removed from the vertex fit, and a variable $x_D$, sensitive to
the invariant mass of the tracks in the jet that have a high
probability of originating from a \bq{} hadron decay. The neural network
output \vnn{} lies between zero and one. Values close to one
indicate a high probability that the vertex is associated with a 
\bq{} hadron decay.

\section{Analysis}

In Section~\ref{textgbbsel} the selection of candidate events for
gluon splitting is described. By changing the $y$-cut and exploiting the
transition from four to three jets, two jets are selected in each
event as candidates to have originated from gluon splitting.
The candidate jets are checked for secondary vertices and
events with two significant secondary vertices are selected.
The event sample is subdivided into two distinct classes, depending on
the event topology. Optimized cuts on the neural network outputs are
applied for each class, to define the candidate events.
In Section~\ref{textgfbsel} a dedicated selection of candidates for the
process $\Zz\to\bbbb$ is discussed, where all four jets are checked
for secondary vertices. Finally, in Section~\ref{textmaxlfit} the
rate of gluon splitting
to \bb{} is calculated. For each of the selected  event samples,
angular distributions sensitive to four-quark final states are studied. 
The rates \gbb{} and \gfb{} are calculated using a binned maximum
likelihood fit. The signal and background selection efficiencies
and these angular distributions are used as input to the fit.

\begin{boldmath}
\subsection{The \protect\qqbb{} event selection}
\label{textgbbsel}
\end{boldmath}

In each four-jet event the $y$-cut is increased, until the event changes to
a three-jet event ($\ycut >\yc{34}$). The two jets that are combined in
this step are considered as candidates for \gtobb{}. The $y$-cut is then
increased further, until the event changes to a two-jet event
($\ycut >\yc{23}$). There are two distinct possibilities for this, as
shown in Figure~\ref{figjetclass}.
\begin{figure}[t]
\begin{center}
\unitlength0.07\picwi
\begin{picture}(12.5,7.5)
\thicklines
\put(1.7,7){\Large\bf class ``2+2''}
\put(5,4){\line(-3,1){3}}\put(5,4){\line(-3,-2){1.5}}
\put(3.5,3){\line(-5,-2){3}}\put(3.5,3){\line(-4,1){3}}
\put(2,5){\line(-2,1){1.5}}\put(2,5){\line(-2,-1){1.5}}
\put(-0.9,5.2){\put(0,0){\gtobb}\put(0,-0.5){candidates}}
\thinlines
\multiput(2,6.5)(0,-0.5){12}{\line(0,-1){0.25}}
\multiput(3.5,6.5)(0,-0.5){12}{\line(0,-1){0.25}}
\put(1,1){\put(-0.3,0.4){4 jets}}\put(2.5,1){\put(-0.3,0.4){3 jets}}
\put(4,1){\put(-0.3,0.4){2 jets}}\put(2,0){\put(-0.2,0.2){\yc{34}}}
\put(3.5,0){\put(-0.2,0.2){\yc{23}}}
\put(2,0){
\thicklines
\put(6.7,7){\Large\bf class ``3+1''}
\put(10,4){\line(-5,-2){4.5}}\put(10,4){\line(-3,1){3}}
\put(8.5,4.5){\line(-3,-1){3}}
\put(7,5){\line(-2,-1){1.5}}\put(7,5){\line(-2,1){1.5}}
\put(4.5,5.2){\put(-0.5,0){\gtobb}\put(-0.5,-0.5){candidates}}
\thinlines
\multiput(7,6.5)(0,-0.5){12}{\line(0,-1){0.25}}
\multiput(8.5,6.5)(0,-0.5){12}{\line(0,-1){0.25}}
\put(6,1){\put(-0.3,0.4){4 jets}}\put(7.5,1){\put(-0.3,0.4){3 jets}}
\put(9,1){\put(-0.3,0.4){2 jets}}\put(7,0){\put(-0.2,0.2){\yc{34}}}
\put(8.5,0){\put(-0.2,0.2){\yc{23}}}
}
\end{picture}
\end{center}
\caption{\label{figjetclass}\it
Illustration of the definition of class ``2+2'' and class ``3+1'' and
the \gtobb{} candidate jet selection.
}
\end{figure}
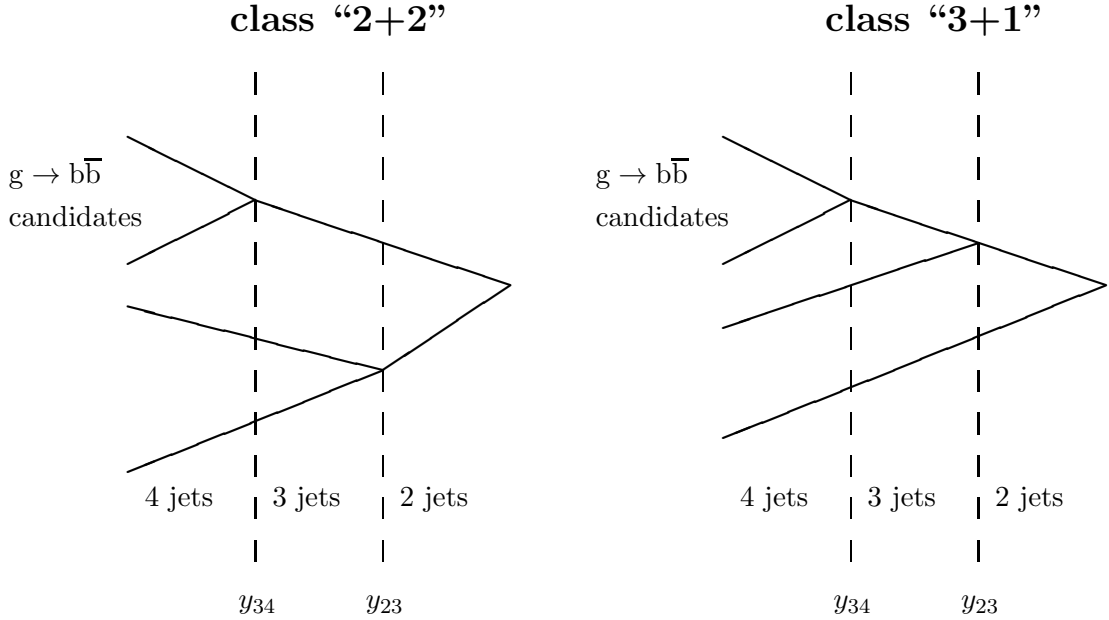
\begin{description}
\item[Class ``2+2'':]
Events belong to this class if none of the original four jets is identical
to any of the two jets obtained after increasing the $y$-cut to force
the event into a two-jet event.
\item[Class ``3+1'':]
Events belong to this class if one of the original four jets is identical
to one of the two jets obtained after increasing the $y$-cut to force
the event into a two-jet event.
\end{description}

Monte Carlo studies show that events from the process \gtobb{} are
preferentially selected in class ``3+1''. The corresponding selection
efficiencies are given in Table~\ref{tablecut}.
For a signal rate $\gbb =\gbbjs\times 10^{-3}$, in class ``3+1'' 
$1.47\%$ of the events are from \gtobb{}.
In class "2+2" only $0.99\%$ of the events are from \gtobb{}.
The event classification procedure selects $\ntwotwo$ data 
events in class ``2+2'' and $\nthreeone$ in class ``3+1''.
The Monte Carlo simulation predicts $\ntwotwomc$  and  $\nthreeonemc$
events, respectively. Implications on the final result for \gbb{} and \gfb{}
from this statistically significant difference in the number of
observed to expected events in those two event classes are considered
in the discussion of the experimental uncertainties.
\begin{table}[t]
\begin{center}
\begin{tabular}{|l||c||c|c|c||c|c|}
\hline
 & \multicolumn{4}{|c||}{Number of events} &
   \multicolumn{2}{|c||}{Selection} \\
\hline
 &      &      & \multicolumn{2}{|c||}{signal} &
 \multicolumn{2}{|c||}{efficiencies} \\
\hline
 & data & bgnd & \qqbb{} & \bbbb{} & $\ebb$ & $\efb$ \\
\hline\hline
\multicolumn{7}{|l|}{The four-jet selection} \\
\hline
 Total four-jet &
 $443334$ & $438191$ & $4263$ & $880$ & $52.72\%$ & $56.95\%$ \\
\hline
 $\quad$ ``2+2'' &
 $235201$ & $238500$ & $1901$ & $373$ & $23.51\%$ & $24.15\%$ \\
 $\quad$ ``3+1'' &
 $208133$ & $199691$ & $2362$ & $507$ & $29.21\%$ & $32.80\%$ \\
\hline\hline
\multicolumn{7}{|l|}{Event selection in Class ``2+2''} \\
\hline
$\quad(l/\sigma)_2>3$ &
 $613$ & $596.2$ & $51.2$ & $17.8$ & $0.63\%$ & $1.15\%$ \\
$\quad\vnn{1}+\vnn{2}>1.7$ &
 $39$ & $25.8$ & $8.6$ & $3.0$ & $0.11\%$ & $0.20\%$ \\
\hline
$\quad$ Sample A &
 $14$ & $6.3$ & $1.2$ & $2.2$ & $0.02\%$ & $0.15\%$ \\
$\quad$ Sample B &
 $25$ & $19.5$ & $7.4$ & $0.8$ & $0.09\%$ & $0.05\%$ \\
\hline\hline
\multicolumn{7}{|l|}{Event selection in Class ``3+1''} \\
\hline
$\quad(l/\sigma)_2>3$ &
 $359$ & $310.6$ & $71.4$ & $22.3$ & $0.88\%$ & $1.31\%$ \\
$\quad\vnn{1}+\vnn{2}>1.1$ &
 $153$ & $92.8$ & $42.9$ & $13.2$ & $0.53\%$ & $0.85\%$ \\
\hline
$\quad$ Sample C &
 $44$ & $40.1$ & $6.7$ & $9.6$ & $0.08\%$ & $0.62\%$ \\
$\quad$ Sample D &
 $109$ & $52.7$ & $36.2$ & $3.6$ & $0.45\%$ & $0.23\%$ \\
\hline\hline
\multicolumn{7}{|l|}{The dedicated \bbbb{} selection} \\
\hline
$\quad(l/\sigma)_3>3$ &
 $628$ & $642.8$ & $18.0$ & $55.8$ & $0.22\%$ & $3.61\%$ \\
$\quad$ remove overlap with A--D &
 $589$ & $604.6$ & $14.0$ & $46.0$ & $0.17\%$ & $2.97\%$ \\
\hline
$\quad$ Sample E &
 $29$ & $21.1$ & $0.3$ & $9.1$ & $0.004\%$ & $0.59\%$ \\
\hline\hline
 {\bf Selected (A--E)} &
 $221$ & $139.7$ & $51.8$ & $25.3$ & $0.64\%$ & $1.64\%$ \\
\hline
\end{tabular}
\end{center}
\caption{\it \label{tablecut}
Events selected at the different steps of the 
analysis (data), and the number of background (bgnd) and signal
events expected from the simulation with
$\gbb=\gbbjs\times 10^{-3}$ and $\gfb=\gfbjs\times 10^{-3}$.
Also shown are the efficiencies to select \qqbb{} $(\mathrm{q=udsc})$
and \bbbb{} signal events, predicted from the Monte Carlo simulation.
}
\end{table}

\begin{figure}[t]
\begin{center}
\epsfig{file=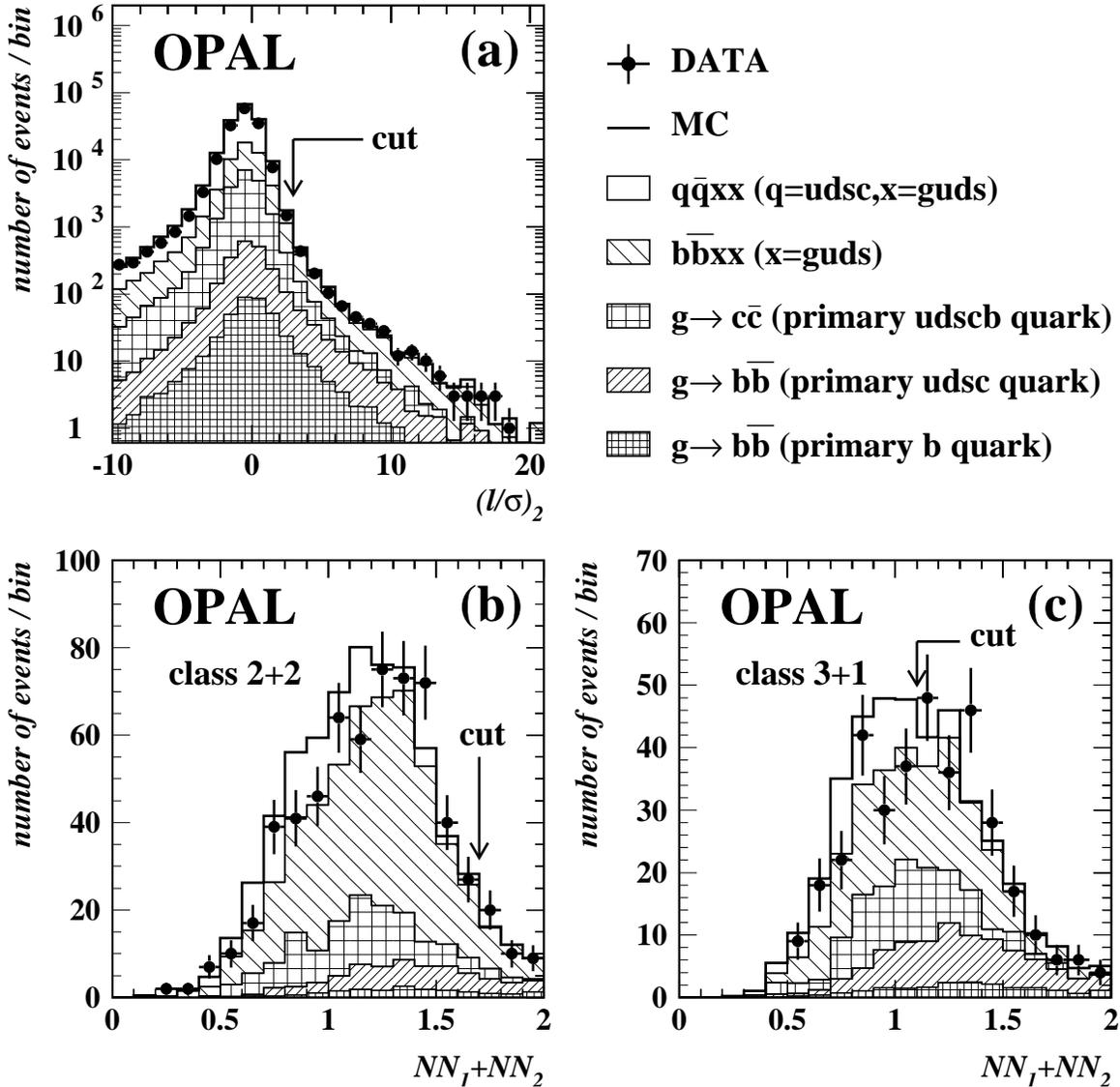,width=\picwi}
\end{center}
\caption{\it\label{figbtwo}
Variables used to select \gtobb{} candidate events. Plot (a) shows the
decay length significance $(l/\sigma)_2$ of the \gtobb{} candidate jet
with the smallest decay length significance. Plot (b) and (c) shows
the sum of the neural network outputs \vnn{1} and \vnn{2} for the
events surviving the $l/\sigma$ cut in class ``2+2'' and ``3+1'', respectively.
The cut values are indicated by arrows. The points show data and the
histogram shows the Monte Carlo simulation. The contributions from
various four-jet processes are indicated.}
\end{figure}

The two \gtobb{} candidate jets selected in the first step are checked
for secondary vertices. If both of these jets have a reconstructed
secondary vertex with $l/\sigma>3$, the event is selected. 
This cut is studied in Fig.~\ref{figbtwo}a, where
the decay length significance $(l/\sigma)_2$ of the second \gtobb{}
candidate jet is shown. The two jets are ordered such that
$(l/\sigma)_1>(l/\sigma)_2$. If the variable $(l/\sigma)_2$ is larger
than 3, the event is selected, otherwise it is rejected. Note that
$(l/\sigma)_2>3$ implies $(l/\sigma)_1>3$. Because only the smaller
decay length significance of the two vertices enters this
distribution, there are many entries at negative $(l/\sigma)_2$
values. The cut $(l/\sigma)_2>3$ reduces the background from
events with light flavors and it also reduces the fraction of events
from the \gtocc{} process.

In the next step, the output of a neural network trained to recognize
vertices from \bq{} hadron decays is calculated for the reconstructed
vertices in the selected events. 
The two \gtobb{} candidates are ordered by their corresponding neural
network  output \vnn{} such that $\vnn{1}>\vnn{2}$.
The sum $\vnn{1}+\vnn{2}$
shows good sensitivity to the signal process \gtobb{} for both event
classes ``2+2'' and ``3+1''. It is depicted in figures \ref{figbtwo}b and
\ref{figbtwo}c.
To enrich the signal, cuts $\vnn{1}+\vnn{2}>1.7$ in class ``2+2'' and
$\vnn{1}+\vnn{2}>1.1$ in class ``3+1'' are made. The cut values are
chosen such as to obtain purities of about $40\%$ for each class of
events. Some discrepancies between data and the prediction are observed in
Fig.~\ref{figbtwo} and in table \ref{tablecut}, mainly at low values of
$l/\sigma$ and low values of $\vnn{1}+\vnn{2}$. These can be
explained by uncertainties in the knowledge of the detector
resolution, studied in Section~\ref{textsysexp}.

\begin{figure}[t]
\begin{center}
\epsfig{file=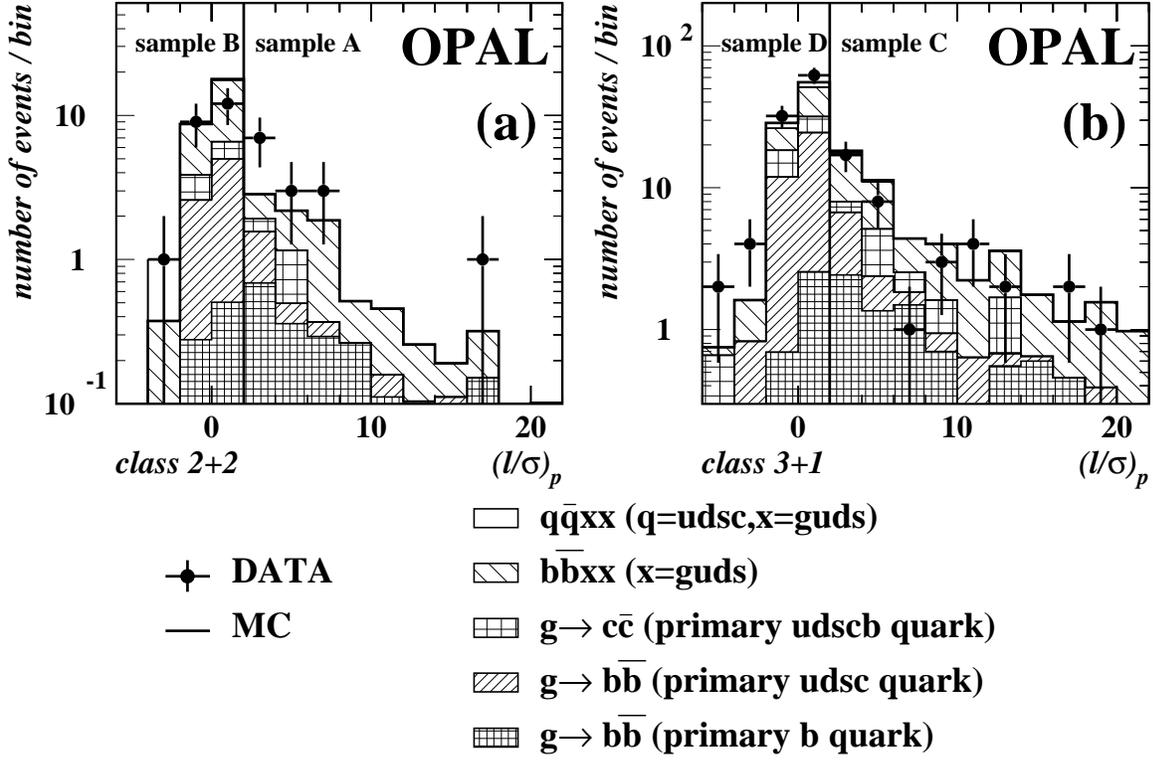,width=\picwi}
\end{center}
\caption{\it\label{figdlsp}
Largest decay length significance $(l/\sigma)_{\mathrm{p}}$ of the
primary quark jet candidates, as defined in the text, after applying all
selection cuts for the \qqbb{} selection. Events from 
class ``2+2'' appear in Fig.~\ref{figdlsp}a, events from class
``3+1'' in Fig.~\ref{figdlsp}b.
The points show data and the
histogram shows the Monte Carlo simulation. The contributions of
various four-jet processes are indicated. The cut at
$(l/\sigma)_{\mathrm{p}}=2$ separating the event samples A from B and C from D is
indicated.}
\end{figure}

Finally, the other two jets that were not considered to be gluon splitting
candidates are checked for secondary vertices to gain some separation
power for the \bbbb{} signal events. These jets are referred to
as ``primary quark jet candidates''. If both of these jets have a
vertex, only the jet with the larger  decay length significance is
considered. Fig.~\ref{figdlsp} shows the
distribution of the decay length significance of this reconstructed
vertex, $(l/\sigma)_{\mathrm{p}}$, for the events selected from class
``2+2'' and ``3+1''.
The variable $(l/\sigma)_{\mathrm{p}}$ shows some separation
between the \bbbb{} events and the \qqbb{} ($\mathrm{q=udsc}$) events.
A cut at $(l/\sigma)_{\mathrm{p}}=2$ is made to select
event samples enriched or depleted with \bbbb{} events. The
following four event samples are defined
\begin{description}
\item[Event sample A:]
Events from class ``2+2'' with $\vnn{1}+\vnn{2}>1.7$ and
$(l/\sigma)_{\mathrm{p}}>2$, enriched in \bbbb{}.
\item[Event sample B:] 
Events from class ``2+2'' with $\vnn{1}+\vnn{2}>1.7$ not selected in
sample~A, depleted in \bbbb{}.
\item[Event sample C:]
Events from class ``3+1'' with $\vnn{1}+\vnn{2}>0.7$ and
$(l/\sigma)_{\mathrm{p}}>2$, enriched in \bbbb{}.
\item[Event sample D:]
Events from class ``3+1'' with $\vnn{1}+\vnn{2}>0.7$ not selected in
sample~C, depleted in \bbbb{}.
\end{description}

\begin{boldmath}
\subsection{The dedicated \protect\bbbb{} event selection}
\label{textgfbsel}
\end{boldmath}

Starting again with the entire sample of four-jet events, selected with
$\yc{34}>0.006$, a dedicated selection of \bbbb{} events is set up,
independent of the cuts presented in the previous section.
To find events with four \bq{} hadrons, events are selected where at
least three decay vertices with a decay length significance
$l/\sigma>3$ are found.
\begin{figure}[t]
\begin{center}
\epsfig{file=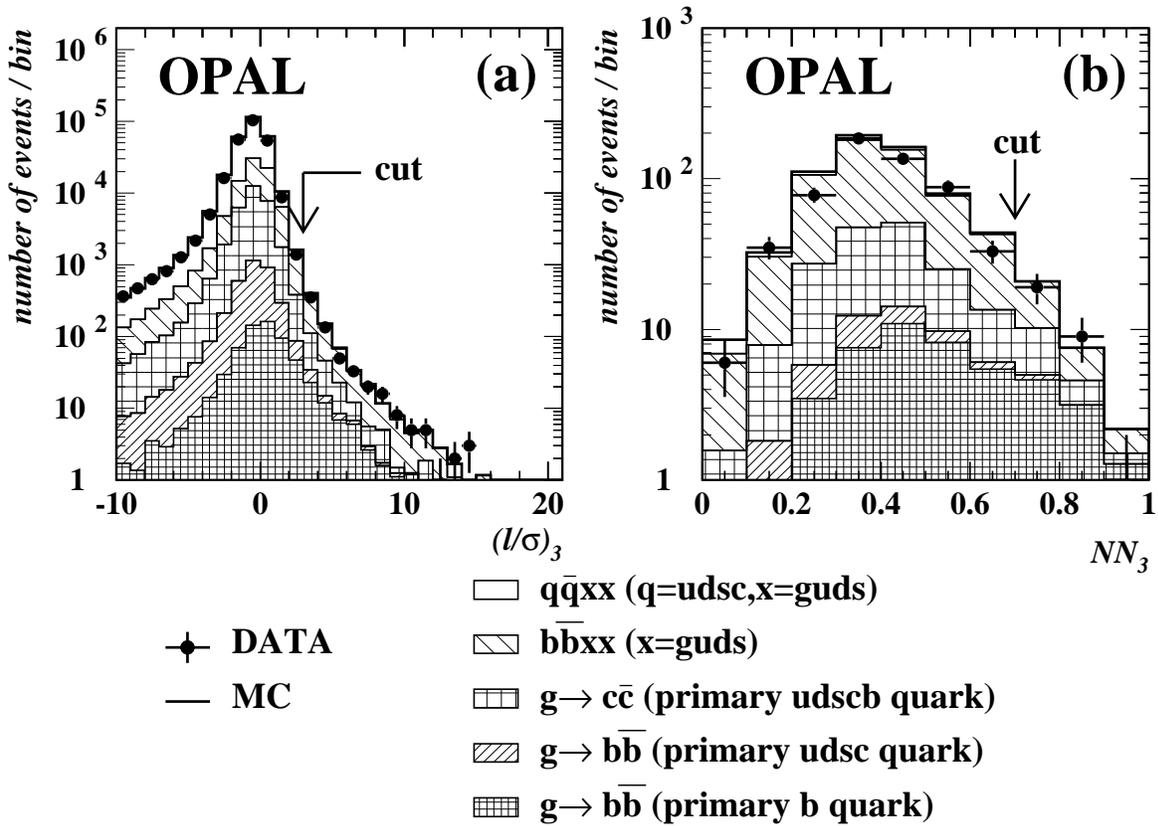,width=\picwi}
\end{center}
\caption{\it\label{figvnn3}
Variables used to select \bbbb{} events.
Plot (a) shows the third largest decay length significance
$(l/\sigma)_3$ of the four jets, plot (b) the third largest neural
network output \vnn{3} with the cut $(l/\sigma)_3>3$ applied and the
overlap with the event samples A-D removed.
The points show data and the
histogram shows the Monte Carlo simulation. The contributions from
various four-jet processes are indicated. Arrows indicate the
position of the cuts.}
\end{figure}
Fig.~\ref{figvnn3}a illustrates this selection of three significant
vertices, showing the third largest decay length significance, denoted
$(l/\sigma)_3$. The \bbbb{} signal is enhanced for large values of this
variable. The cut $(l/\sigma)_3>3$ suppresses light flavors, most of
the \gtocc{} events, and also the \qqbb{} ($\mathrm{q=udsc}$) events.
Fig.~\ref{figvnn3}b shows the third largest neural network output
\vnn{3} for all selected vertices, where the events already selected
in sample A--D are excluded to avoid double counting. The background
dominates the region of low \vnn{3}, while the \bbbb{}
signal extends to high values of \vnn{3}.
A cut $\vnn{3}>0.7$ is chosen to select the final \bbbb{} candidates
from distribution, denoted as sample E.
\\

\subsection{\protect\begin{boldmath}
Calculation of \protect\gbb{} and \protect\gfb{}\protect\end{boldmath}}
\label{textmaxlfit}

After applying all cuts, \ncand{} events remain in the event samples
A--E, where the simulation with $\gbb=\gbbjs\times 10^{-3}$ and
$\gfb=\gfbjs\times 10^{-3}$ predicts $\nexpect\pm\dnexpect$ events.
The efficiencies for selecting signal events, \qqbb{}
($\mathrm{q=udsc}$) or \bbbb{} events, are denoted \ebb{} and
\efb{}. The total efficiencies as obtained from the simulation are
$\ebb=(\effbb\pm\deffbb)\%$ and $\efb=(\efffb\pm\defffb)\%$. The
uncertainties quoted on the number of expected events and the
selection efficiencies are due to the limited number of Monte Carlo events.
Table~\ref{tablecut} summarizes the number of selected events for the
cuts applied, the number of signal and background events expected from
the simulation, and the efficiencies to select signal or background
reactions.

\begin{figure}[t]
\begin{center}
\epsfig{file=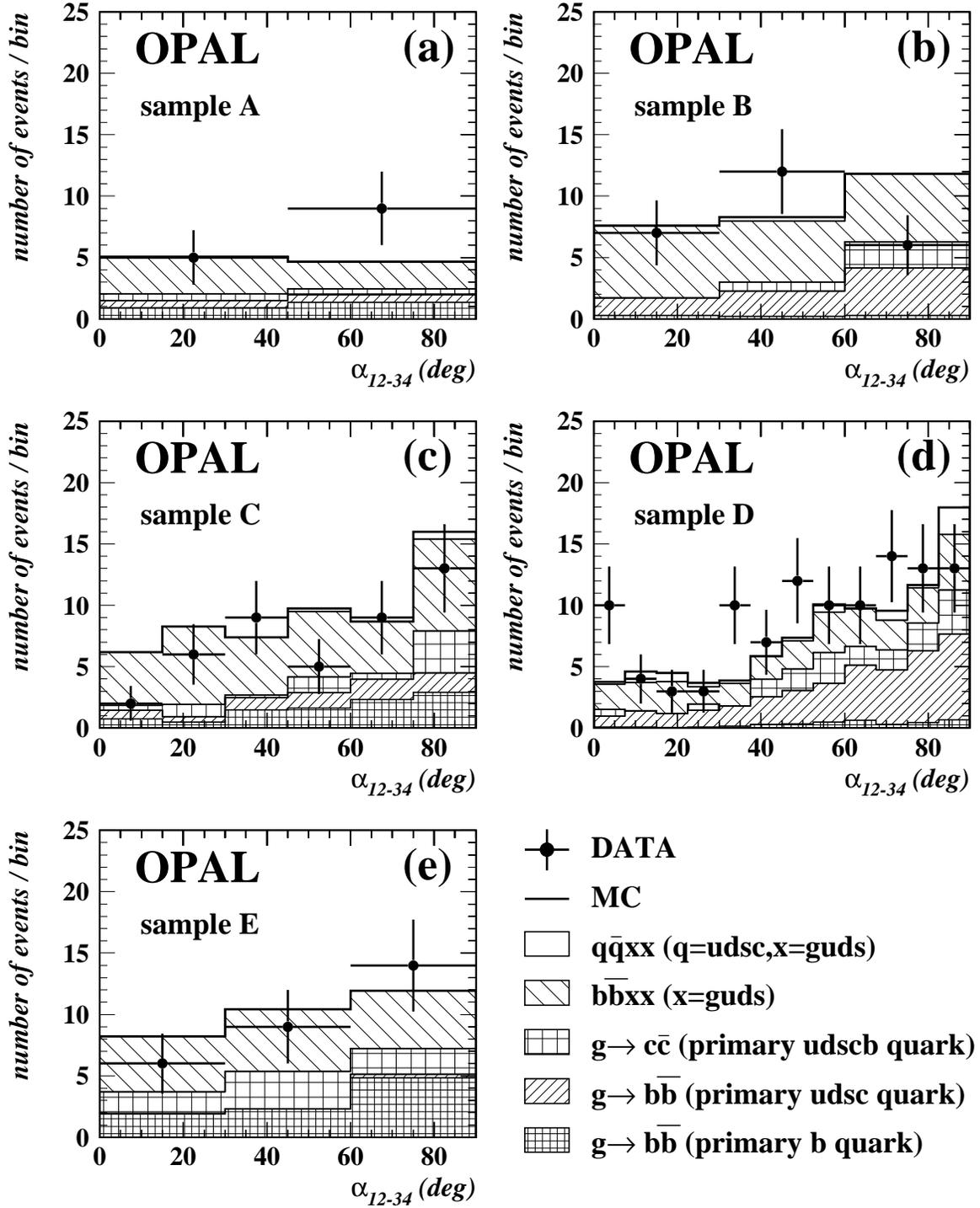,width=\picwi}
\end{center}
\caption{\it\label{figbz}
The angle \bz{} between the plane formed by the two \gtobb{} candidate
jets and the plane formed by the other two jets for the
\gtobb{} candidate events after applying all cuts. Figures
a--e correspond to the event samples A--E. The
data are shown as full points with error bars, the Monte Carlo
simulation as a solid line. The contributions from various four-jet
processes are indicated.}
\end{figure}

For the \ncand{} signal events the angle \bz{} between two jet-jet
planes is studied to distinguish signal events from the background, dominated
by events with two quarks and two gluons. The
first plane is spanned by the two jets that are joined into one jet by
the jet-algorithm at the transition from four to three jets. The other
plane is formed by the other two jets. The
definition of this angle \bz{} is similar to the angular correlation
proposed in \cite{bzerwas} to measure the QCD color factors. 
In Fig.~\ref{figbz}, the \bz{}
distribution is shown to be consistent with the theoretical prediction. 
The signal appears preferentially at high values of \bz{} while the
background has a flatter distribution.

A maximum likelihood fit of the \nbins{} bins in \bz{} shown in
Fig.~\ref{figbz} is performed to extract \gbb{} and \gfb{},
assuming Poisson distributions calculated from the signal and
background efficiencies for each bin. The likelihood function is given by
\[
 -\ln{\cal L} = \sum_{i=1}^{\nbins} ( \mu_i - d_i \ln \mu_i ) +
\mathrm{const},
\]
where $d_i$ is the number of observed events in bin $i$ and $\mu_i$ is
the corresponding number of predicted events (assumed to be the mean
of the Poisson distributions). The latter is given by
\[
 \mu_i = N_{\mathrm{4-jet}} \frac{(1-\gcc-\gbb)\eqq^i + \gcc \ecc^i 
     + (\gbb -\gfb) \ebb^i + \gfb \efb^i}
  {(1-\gcc-\gbb)\eqq^{\mathrm{4-jet}} + \gcc \ecc^{\mathrm{4-jet}}
  +(\gbb -\gfb) \ebb^{\mathrm{4-jet}} + \gfb \efb^{\mathrm{4-jet}}},
\]
where $N_{\mathrm{4-jet}}$ is the total number of selected four-jet
events used for this analysis.
The rate of gluon splitting to \cc{} per hadronic \Zz{}
decay is denoted \gcc{}.
The two rates \gbb{} and \gfb{} are taken as the fit parameters.
The probabilities to select the signal process for primary light quarks
and the \bbbb{} events in bin $i$ are denoted
$\ebb^i$ and $\efb^i$, while
$\ecc^i$ and $\eqq^i$ are the probabilities to select the
\gtocc{} events or other background events in bin $i$. Finally
$\ebb^{\mathrm{4-jet}}$, $\efb^{\mathrm{4-jet}}$, $\ecc^{\mathrm{4-jet}}$,
$\eqq^{\mathrm{4-jet}}$ are the efficiencies for events in the signal
and the background channels to pass the four-jet selection.
The likelihood analysis leads to the result
\begin{eqnarray*}
\gbb & = & (\ngbb\pm \dngbb)\times 10^{-3},\\
\gfb & = & (\ngfb\pm \dngfb)\times 10^{-3},
\end{eqnarray*}
with a correlation coefficient between \gbb{} and \gfb{} of \gfgbcorr{}.
The errors are statistical only. 

\section{Evaluation of systematic uncertainties}

Various sources of systematic uncertainties are considered, as
summarized in Table~\ref{tablesys} and discussed in the following.
Additional cross-checks are presented in Section~\ref{textxcheck}.
\begin{table}[t]
\begin{center}
\begin{tabular}{|l||c|c|c|}
\hline
Source of systematic error & $\Delta\gbb\times 10^3$ & 
                             $\Delta\gfb\times 10^3$ & Correlation \\
\hline\hline
Model dependence &        $\dmodel$ & $\dfmodel$ & \cmodel \\
\bq{} quark mass &        $\dbmass$ & $\dfbmass$ & \cbmass \\
\hline
Monte Carlo statistics &  $\dmcstat$ & $\dfmcstat$ & \cmcstat \\
Detector simulation &     $\dtracking$ & $\dftracking$ & \ctracking  \\
Event classification &    $\dnorm$ & $\dfnorm$ & \cnorm \\
\hline
Flavor composition \Rbfour{} &  $\dgbbrb$ & $\dfgbbrb$ & \cgbbrb \\
Gluon splitting to charm \gcc{} &  $\dgbbgcc$ & $\dfgbbgcc$ & \cgbbgcc \\
\hline
Bottom fragmentation        &  $\dbfrag$ & $\dfbfrag$ & \cbfrag \\
Bottom decay multiplicities &  $\dbmult$ & $\dfbmult$ & \cbmult \\
Bottom production rates     &  $\dbprod$ & $\dfbprod$ & \cbprod \\
Bottom hadron lifetimes     &  $\dbtau$  & $\dfbtau$ & \cbtau \\
\hline
Charm fragmentation         &  $\dcfrag$ & $\dfcfrag$ & \ccfrag \\
Charm decay multiplicities  &  $\dcmult$ & $\dfcmult$ & \ccmult \\
Charm production rates      &  $\dcprod$ & $\dfcprod$ & \ccprod \\
Charm hadron lifetimes      &  $\dctau$  & $\dfctau$   & \cctau \\
\hline\hline
Total systematic error &
 \begin{boldmath}$\dngbbsys$\end{boldmath} &
 \begin{boldmath}$\dngfbsys$\end{boldmath} &
 \begin{boldmath}$\gfgbcorrsys$\end{boldmath} \\
\hline
\end{tabular}
\end{center}
\caption{\it \label{tablesys}
Summary of the systematic uncertainties on the rates \gbb{} and \gfb{}
and the corresponding correlations}
\end{table}

\subsection{Model dependence}
\label{textmodel}

Though the parton shower approach as implemented in \Jetset{}
is expected to describe the gluon splitting process quite well
\cite{seymour95}, it is desirable to look at various alternative models and at
exact calculations.
In \Newpythia{} \cite{pythia99}, using the option {\tt MSTJ(42)=3},
the calculation of the opening angle in the gluon splitting
process has been modified to take mass effects into account, as
compared to \Jetset{}. The \Newpythia{} prediction for the \gtobb{}
process is used to evaluate the main results of this analysis.
The \Herwig{} \cite{herwig} Monte Carlo
generator\footnote{This version was already used in \cite{aleph}.}
provides an alternative model for the parton shower.
Another approach is given by the Color Dipole Model, using the event
generator \Ariadne{} \cite{ariadne}.

The program \WPHACTnv{} \cite{accomando97}
implements \epem{} annihilation to four fermions,
where the masses of \bq{} quarks are taken into account for the
calculation of the matrix elements. It can
be used to study \bbbb{} and \bb\qq{} final states. Version 1.3 of
\WPHACTnv\footnote{Provided to us by courtesy of the author.}
\cite{ballestrero99} allows for the possibility to
switch off Feynman graphs where the \bq{} quarks couple to the \Zz{} or
$\gamma$, enabling studies of the gluon splitting
process at tree level.
For the case of primary light quarks, calculations are available
including next-to-leading order logarithmic terms 
\cite{seymour95,miller98}. The shapes of the
\WPHACT{} predictions are in agreement with these calculations in the
regions where such a comparison is possible.

\begin{figure}[t]
\begin{center}
\epsfig{file=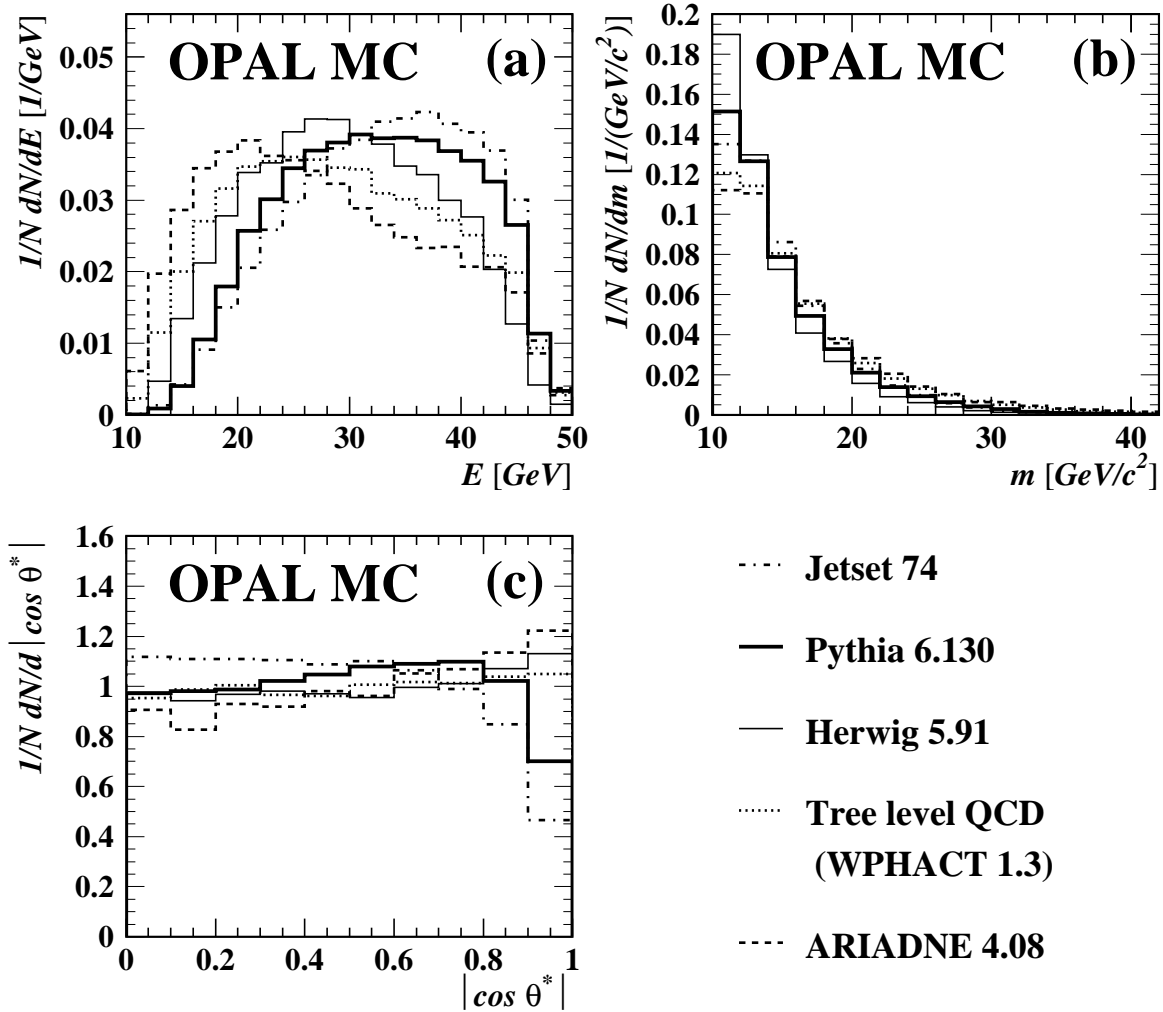,width=\picwi}
\end{center}
\caption{\it\label{figgluon} Normalized distributions of three kinematic
variables of the gluon that splits to \bb{} for different models and
calculations, as described in the text: (a) the energy $E$ , (b) the 
virtuality $m$, (c) the decay angle $\vert\cos\theta^\star\vert$ of the \bb{}
pair in the gluon rest frame.
}
\end{figure}

Fig.~\ref{figgluon} shows differential distributions of
kinematic variables of the gluon that splits to
\bb{}, calculated with \Newpythia{},\Jetset{}, \Herwig{}, \Ariadne{} and
\WPHACT{}. 
The ``gluon'' variables have been calculated from the \bb{} quark pair
at the end of the shower. For the case of \bbbb{} events the \bb{}
quark pair with the lowest invariant mass is chosen. The hardest energy
spectrum is predicted by \Jetset{}, while \Ariadne{} leads to the
softest energy spectrum, when comparing the five models. For the gluon
virtuality the \Ariadne{} prediction leads to the hardest
distribution, while \Herwig{} predicts the softest spectrum. For the
decay angle in the gluon rest frame the
models differ most significantly at high
$\vert\cos\theta^\star\vert$. The extreme cases are covered by
\Jetset{} with a low number of events in this region and by \Ariadne{}
showing an enhancement of events with increasing
$\vert\cos\theta^\star\vert$. Note that the \Newpythia{} prediction is
always well between those extreme cases, this is why we decided to use 
it for the main analysis results.
 
To study hadronisation and detector effects, we use two sets of events
generated either with \Jetset{} or with \Newpythia{}, and having a detailed
hadronisation and detector simulation. The \Jetset{} sample is only
used for systematic checks.

\begin{figure}[t]
\begin{center}
\epsfig{file=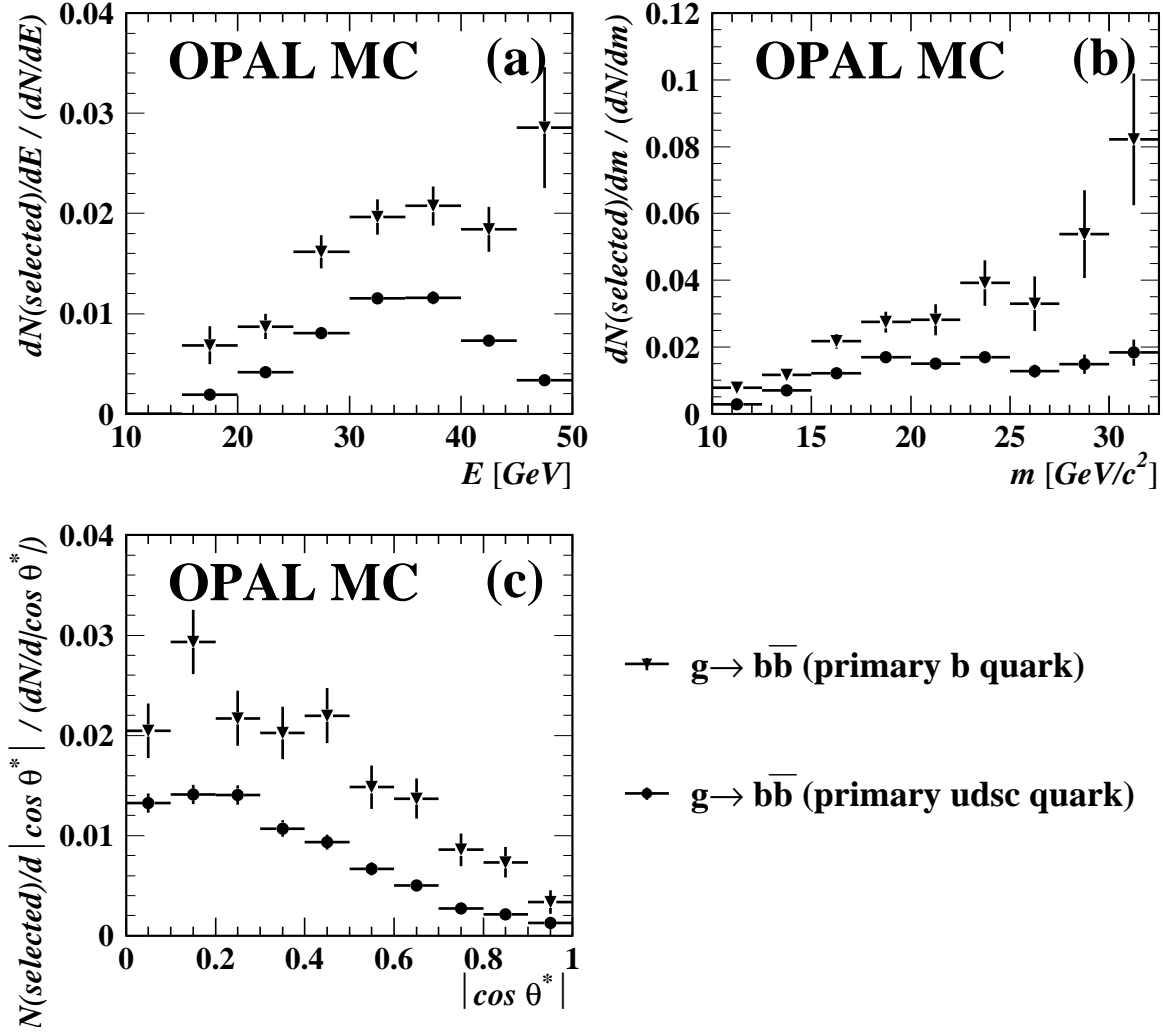,width=\picwi}
\end{center}
\caption{\it\label{figeff} Selection efficiencies as a function
  of kinematic variables of the gluon that splits to \bb{}: (a) the
  energy $E$, (b) the virtuality $m$, (c) the decay angle
$\vert\cos\theta^\star\vert$ of the \bb{} pair in the gluon rest-frame. 
  The efficiencies to select \qqbb{} $(\mathrm{q=udsc})$
and \bbbb{} events are indicated.
}
\end{figure}

The differential efficiencies to select the signal events,
\qqbb{} ($\mathrm{q=udsc}$) and \bbbb{} in this analysis, are shown in figure
\ref{figeff}. They are evaluated with the \Newpythia{} event sample.
The efficiencies are low for small gluon energies $E$  and gluon
virtualities $m$, as it is difficult to detect the two \bq{} hadrons
in two separate jets in this region. The same is true for decay angles
$\vert\cos\theta^{\star}\vert$ close to one, because in this case the two
\bq{} hadrons have a small transverse momentum relative to the gluon
flight direction.

To evaluate model-dependent effects, the complete analysis chain is
performed using either the \Newpythia{} or the \Jetset{} event
sets. In addition, the \Newpythia{} events are
weighted such as to reproduce the gluon distributions from
Fig.~\ref{figgluon} predicted by \Herwig{}, \Ariadne{} and \WPHACT{},
and the analysis is repeated.
Table~\ref{tablemodel} summarizes the individual results of
these studies.
\begin{table}[t]
\begin{center}
\begin{tabular}{|l|l|c|c|}
\hline
Model & Implementation & $\gbb\times 10^3$ & $\gfb\times 10^3$ \\
\hline
\Jetset{}     & Full detector simulation &
 $\ngbbjs\pm\dngbbjs$ & $\ngfbjs\pm\dngfbjs$ \\
\Newpythia{} (default) & Full detector simulation & 
 $\ngbb\pm\dngbb$ & $\ngfb\pm\dngfb$ \\
\WPHACT{}     & Event reweighting          &
 $\ngbbwp\pm\dngbbwp$ & $\ngfbwp\pm\dngfbwp$ \\
\Herwig{}     & Event reweighting          &
 $\ngbbhw\pm\dngbbhw$ & $\ngfbhw\pm\dngfbhw$ \\
\Ariadne{}     & Event reweighting          &
 $\ngbbar\pm\dngbbar$ & $\ngfbar\pm\dngfbar$ \\
\hline
\end{tabular}
\end{center}
\caption{\it\label{tablemodel}Results obtained for various Monte Carlo models.}
\end{table}
To study possible deficiencies in the reweighting procedure, it is 
repeated based on the \Jetset{} event sample rather than on the
\Newpythia{} event sample. This results in slightly lower values of
\gbb{} and larger values of \gfb{}, as compared to the \Herwig{},
\Ariadne{} and \WPHACT{} values shown in table \ref{tablemodel}. The
most significant differences to the standard reweighting procedure are
observed for \Herwig{}, namely  $\gbb=2.88\times10^{-3}$ and 
$\gfb=0.53\times10^{-3}$.

Finally, the largest differences on \gbb{} and \gfb{} to the central value
are taken as systematic uncertainty due to limited knowledge of the
gluon splitting mechanism. For \gbb{} this is the \Newpythia{}--\Jetset{}
difference, while for \gfb{} this is the \Newpythia{}--\Herwig{}
difference, where the \Herwig{} result is calculated with the help of
reweighted \Jetset{} events.

\begin{boldmath}
\subsection{Dependence on the $\protect\bq$ quark mass}
\label{textbmass}
\end{boldmath}

The influence of the \bq{} quark mass assumed for the Monte Carlo simulation 
has been studied by changing the \bq{} quark mass in the \Newpythia{}
program\footnote{The relevant parameter in the \Newpythia{} program is
{\tt PMAS(5,1)}.} and investigating the properties of the \bq{}
hadrons produced in the fragmentation process. The central value for
the \bq{} quark mass in \Newpythia{} used throughout this analysis is
$5\,\gevm$. The shapes of the distributions of various kinematic variables
are used to reweight the \gtobb{} Monte Carlo events corresponding to a \bq{}
quark mass of $4.5\,\gevm$ and $5.25\,\gevm$. This
variation covers the uncertainty of the \bq{} quark pole mass
\cite{pdg} up to the B meson mass. Note that the way the quark mass
parameter is used in \Newpythia{} corresponds rather to a
constituent quark mass than to a pole mass definition of the quark
mass \cite{sjostrandmb}, which justifies the choice of $5\,\gevm$ for
the central value.

The quantities used for the reweighting process are 
the momenta of the two \bq{} hadrons produced in the fragmentation
process and their invariant mass.
These variables are chosen because the efficiency to identify \bq{}
hadrons strongly depends on their momentum. The
efficiency to resolve the two \bq{} hadrons in different jets depends
on their invariant mass.
The larger deviations from the standard results are found for a \bq{}
quark mass of $4.5\,\gevm$. They are taken into account as 
systematic uncertainties.
\\

\subsection{Experimental sources of systematic uncertainties}
\label{textsysexp}

The modeling of the OPAL detector is important 
because the analysis depends on an accurate understanding of
the decay vertex reconstruction.
The \bq{} tagging efficiencies are
mainly sensitive to the parameters modeling the production and
decay of \bq{} hadrons. This will be discussed in Section~\ref{textsyshf}.
The light quark tagging efficiencies in the Monte Carlo simulation
are mainly sensitive to details in the modeling of the tracking system.
Studies are done by increasing the difference of the reconstructed
track parameters with respect to the true track parameters in the
Monte Carlo simulation by $10\%$ to cover uncertainties in the
knowledge of the flavor tagging variables \cite{opalrb98}.
This smearing is applied separately
for parameters defined in the $(r,\phi)$ and the $(r,z)$ plane. In
$(r,\phi)$ a simultaneous smearing of the distance 
of closest approach, $d_0$, and the azimuthal angle $\phi_0$ is
performed. In $(r,z)$ smearing is done simultaneously for the $z$ coordinate of
the point of closest approach in $(r,\phi)$ and the polar angle $\theta$.
Finally the gluon splitting analysis is repeated using the modified
Monte Carlo sets. The uncertainties from the smearing in $(r,\phi)$
and $(r,z)$ are added quadratically.

The normalization to the number of four-jet events revealed
differences in the population of the event classes ``2+2''
and ``3+1''when comparing the data with the Monte Carlo prediction,
as discussed in Section~\ref{textgbbsel}. This is addressed by
repeating the likelihood fit, using the numbers of events and selection
efficiencies of the ``2+2'' and ``3+1'' classes, rather than the
number of four-jet events with their corresponding selection efficiencies.
The variation found is assigned as systematic uncertainty due
to the event classification.

\subsection{Uncertainties from heavy flavor physics}
\label{textsyshf}

The rate of primary \bb{} production in the
four-jet sample \Rbfour{} can be measured using a double-tag
technique. Agreement with the Monte Carlo simulation within $\devrb$
is found, where the statistical uncertainties are in the order of $1.5\%$.
The rate \Rbfour{} in the Monte Carlo sample has been changed by
$\devrb$ and the differences in \gbb{} and \gfb{} to our standard
results are taken as systematic uncertainties. Varying the rate of
primary \cc{} production \Rc{} within its uncertainties only has a negligible
effect on the results. 

The rate of secondary \cc{} quark pairs from gluons, \gcc{}, has been
measured by OPAL to be $\gcc =(\gccval\pm\dgccval)\times 10^{-2}$
\cite{gtocc}. This measurement is based on lepton
identification and the reconstruction of $\mathrm{D}^\star$ mesons in
three-jet events. It can be considered as statistically independent of
the \gbb{} measurement presented here, which is based on lifetime-tags in
four-jet events. The rate \gcc{} thus is varied within its uncertainty
to obtain systematic uncertainties on \gbb{} and \gfb{}.
The explicit dependence of the results of this analysis on \gcc{} is given by
\[
  \frac{\dgbb}{\gbb}= \delrgbbrgcc\times\frac{\Delta\gcc}{\gcc}, \quad
  \frac{\dgfb}{\gfb}= \delrgfbrgcc\times\frac{\Delta\gcc}{\gcc}.
\]

The fragmentation functions for charm and bottom quarks
are varied to reflect the uncertainties in the knowledge of the
average scaled energies for D and B mesons, ${\langle
x_E\rangle}_{\cq}=0.484\pm 0.008$ and ${\langle
x_E\rangle}_{\bq}=0.702\pm 0.008$ \cite{lephfwg}. This
is done by varying the parameter $\epsilon$ in the
parameterization of the fragmentation function suggested by Peterson
et.~al.~\cite{peterson}. In addition the sensitivity to the shape of
the fragmentation function is checked, using the models suggested
by Collins and Spiller \cite{collins} and Kartvelishvili
\cite{kartvel}. The parameters of these models are chosen to reproduce
the measured values of the mean scaled energies of the D and B mesons
in the Monte Carlo simulation, and the larger deviations in \gbb{} and
\gfb{} are used as systematic uncertainties from this source.
The uncertainties from the knowledge of the mean
scaled energies and the shape of the distributions are added
in quadrature.

The charged decay multiplicities of the D mesons are varied within
the errors given in \cite{markiii} around the \Jetset{} prediction.
Particles from decay-chains of \bq{} hadrons are
excluded from this variation. The mean charged decay multiplicity of
weakly decaying \bq{} hadrons is varied within  $n_B=4.995\pm 0.062$
\cite{lephfwg}. This multiplicity includes secondary decays of charm hadrons
produced in the decay chains. Variations of the neutral decay
multiplicities have not been studied, but are expected to be small.

The production rates of charm and bottom hadrons
in the Monte Carlo simulation are varied within the ranges given by
the LEP Electroweak Working group  \cite{lephfwg} and 
the Particle Data Group \cite{pdg}, respectively. This variation
includes particles from primary \bb{} and \cc{} production as well as
those from gluon splitting to heavy quarks.

The lifetimes of charm and bottom hadrons are varied around
their central value according to the numbers given by the Particle Data Group
\cite{pdg}.

\subsection{Results}
All contributions to the systematic uncertainties are added in
quadrature, leading to the results
\begin{eqnarray*}
\gbb & = & (\ngbb\pm \dngbb \mathrm{(stat.)}\pm
      \dngbbsys \mathrm{(syst)})\times 10^{-3}, \\
\gfb & = & (\ngfb\pm \dngfb \mathrm{(stat.)}\pm
      \dngfbsys \mathrm{(syst)})\times 10^{-3}.
\end{eqnarray*}
The systematic errors on \gbb{} and \gfb{} have a correlation
$\gfgbcorrsys$. This large positive correlation can be understood from
the fact that most of the systematic variations influence the selection
efficiencies for both the \gtobb{} and the \bbbb{} signal in the same
direction.

\section{Additional cross-checks}
\label{textxcheck}

\begin{figure}[t]
\begin{center}
\epsfig{file=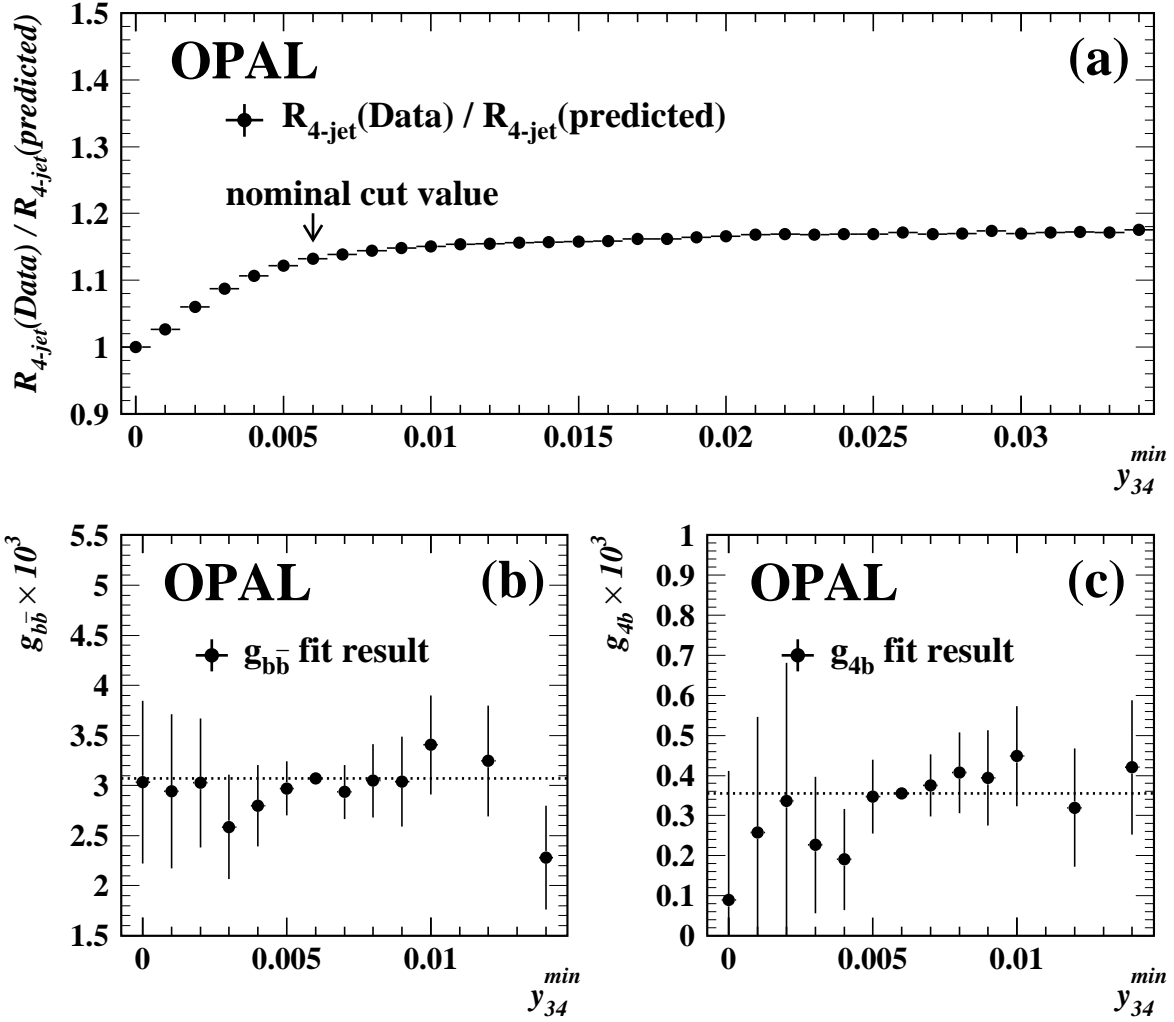,width=\picwi}
\end{center}
\caption{\it\label{figycut}
Dependence of the four-jet rate $R_{\mathrm{4-jet}}$ and the fit results on
$\yc{34}^{\mathrm{min}}$, the lower cut in the jet resolution
parameter \yc{34}. Figure (a) shows the ratio of the number of events
in the data by the number of events predicted from the Monte Carlo
simulation, figures (b) and (c) show the fit result for \gbb{} and
\gfb{}. The error bars in (b) and (c) correspond to the statistical
errors independent from the central value
$\yc{34}^{\mathrm{min}}=0.006$.}
\end{figure}

The four-jet selection is defined using a cut
$\yc{34}>\yc{34}^{\mathrm{min}}$. Fig.~\ref{figycut}a
shows the number of events in the data divided by the predicted number
of events as a function of this cut.
The disagreement in the rate of
four-jet events, as discussed in Section~\ref{textfourjet}, 
is clearly visible. The fit results for \gbb{} and
\gfb{} as a function of this cut are studied in Fig.~\ref{figycut}b
and  \ref{figycut}c. The fit results are stable under the variation of
$\yc{34}^{\mathrm{min}}$ within the independent statistical
error. Note that the variation of $\yc{34}^{\mathrm{min}}$ from $0$ to
$0.014$ corresponds to a variation in the number of candidate events
from $1003$ to $93$ and a variation of the estimated \gtobb{} signal
purity from $10\%$ to $50\%$.
These checks show that the treatment of the normalization to the
four-jet rate in the likelihood fit is correct.

To check the fitting procedure a likelihood fit is applied to the normalized
signal and background shapes in \bz{}, calculating directly the
fractions of \qqbb{} ($\mathrm{q=udsc}$) and \bbbb{} events in the final event
selection from these shapes. Using the number of selected events, the
total number of hadronic \Zz{} decays and the signal efficiencies,
\gbb{} and \gfb{} can be calculated from these event fractions. This
procedure avoids the need to know the absolute selection
efficiency for the background. Alternatively, the likelihood fit is
repeated using only the individual signal and background selection
efficiencies of  the event samples A--E. This analysis is thus
independent of the variable \bz{}. Both results are compatible with
the default method.

The \kT{} algorithm is used in this analysis not only to define
the jets, but also to define the event topology. Therefore it is
important to test the analysis, using a different algorithm. The
\ezero{} \cite{jade} jet clustering scheme is used as an
alternative method. 
Four-jet events are selected with a cut $y_{34}^{\mathrm JADE}>0.015$, 
resulting in
a similar number of four-jet events as compared to the default analysis
which is based on the \kT{} algorithm. To classify the events and
define the gluon splitting candidates, as described in section
\ref{textgbbsel}, and to define the angle \bz{}, as described in
Section~\ref{textmaxlfit}, the new definition of the jet resolution
parameter in the \ezero{} scheme is used. All other analysis cuts
are unchanged with respect to the standard analysis.
This results in \ncanjade{} selected events. The result of the
likelihood fit using the \ezero{} algorithm is
$\gbb^{\mathrm{JADE}}=(\ngbbjade\pm \dngbbjade)\times 10^{-3}$ and 
$\gfb^{\mathrm{JADE}}=(\ngfbjade\pm \dngfbjade)\times 10^{-3}$ 
(statistical uncertainties only).
These results are compatible with the results obtained with the
\kT{} algorithm within the statistical errors, considering the
small fraction of events common to both selections.

\section{Summary}
A measurement of the inclusive rate of gluon splitting to \bb{} per
hadronic \Zz{} decay has been performed using data taken by
OPAL. The result is 
\[
\gbb= (\ngbb\pm \dngbb \pm\dngbbsys
)\times 10^{-3}.
\]
The rate of events with four bottom quarks is measured simultaneously, 
with the result
\[
\gfb = (\ngfb\pm \dngfb \pm\dngfbsys
)\times 10^{-3}.
\]
The correlation of \gbb{} and \gfb{} is \gfgbcorr{} from 
statistical uncertainties and \gfgbcorrsys{} from systematic
uncertainties. The ratio $\gfb{}/\gbb{}$ is thus measured to be
\[
\frac{\gfb}{\gbb}=\fracfb \pm \dfracfb\pm \dfracfbsys
.
\]
This is in agreement with the simplified expectation
$\gfb{}/\gbb{}\approx\Rb$ and with a more detailed calculation
performed by DELPHI \cite{delphifour}, predicting $\gbb /\gfb =0.1833\pm
0.0003$.
The results for \gbb{} and \gfb{} are compatible with previous
measurements \cite{delphi,aleph,delphifour}
as well as theoretical predictions \cite{seymour95,frixione97}.

\pagebreak

\appendix
\section*{Acknowledgments}

We particularly wish to thank the SL Division for the efficient operation
of the LEP accelerator at all energies
 and for their continuing close cooperation with
our experimental group.  We thank our colleagues from CEA, DAPNIA/SPP,
CE-Saclay for their efforts over the years on the time-of-flight and trigger
systems which we continue to use.  In addition to the support staff at our own
institutions we are pleased to acknowledge the  \\
Department of Energy, USA, \\
National Science Foundation, USA, \\
Particle Physics and Astronomy Research Council, UK, \\
Natural Sciences and Engineering Research Council, Canada, \\
Israel Science Foundation, administered by the Israel
Academy of Science and Humanities, \\
Minerva Gesellschaft, \\
Benoziyo Center for High Energy Physics,\\
Japanese Ministry of Education, Science and Culture (the
Monbusho) and a grant under the Monbusho International
Science Research Program,\\
Japanese Society for the Promotion of Science (JSPS),\\
German Israeli Bi-national Science Foundation (GIF), \\
Bundesministerium f\"ur Bildung und Forschung, Germany, \\
National Research Council of Canada, \\
Research Corporation, USA,\\
Hungarian Foundation for Scientific Research, OTKA T-029328, 
T023793 and OTKA F-023259.\\


\begin{thebibliography}{87}

\bibitem{seymour95}
M.~H.~Seymour, Nucl.~Phys.~{\bf  B436} (1995) 163-183.

\bibitem{miller98}
D.~J.~Miller and M.~H.~Seymour, Phys.~Lett.~{\bf B435} (1998) 213.

\bibitem{jetset}
T.~Sj\"ostrand, Comp.~Phys.~Comm.~{\bf 39} (1986) 347;
\\ T.~Sj\"ostrand, Comp.~Phys.~Comm.~{\bf 82} (1994) 74.

\bibitem{frixione97}
S.~Frixione {\em et al.}, CERN-TH-97-16, publ.~in: Heavy Flavours~2,
A.~J.~Buras (ed.) and M.~Lindner (ed.), Singapore World Sci.~(1998).

\bibitem{kniehl}
B.~A.~Kniehl and J.~H.~K\"uhn, Nucl.~Phys.~{\bf B329} (1990) 547-573.

\bibitem{delphi}
DELPHI Collaboration, P.~Abreu {\em et al.}, Phys.~Lett.~{\bf B405}
(1997) 202.

\bibitem{aleph}
ALEPH Collaboration, R.~Barate {\em et al.}, Phys.~Lett.~{\bf B434} 
(1998) 437.

\bibitem{delphifour}
DELPHI Collaboration, P.~Abreu {\em et al.}, Phys.~Lett.~{\bf B462}
(1999) 425.

\bibitem{accomando97}
E.~Accomando and A.~Ballestrero, Comp.~Phys.~Comm.~{\bf 99} (1997) 270.

\bibitem{bzerwas}
M.~Bengtsson and P.~Zerwas, Phys.~Lett.~{\bf B208} (1988) 306.

\bibitem{opaldet}
OPAL Collaboration, K.~Ahmet {\em et al.}, Nucl.~Instr.~Meth.~{\bf
A305} (1991) 275.

\bibitem{opalsi}
P.~P.~Allport {\em et al.}, Nucl.~Instr.~and Meth.~{\bf A324} (1993) 34.

\bibitem{opalsi1}
P.~P.~Allport {\em et al.}, Nucl.~Instr.~and Meth.~{\bf A346} (1994)
476.

\bibitem{opalsi2}
S.~Anderson {\em et al.}, Nucl.~Instr.~and Meth.~{\bf A403} (1998) 326.

\bibitem{rmhsel}
OPAL Collaboration, G.~Alexander {\em et al.}, Z.~Phys.~{\bf C52} (1991) 175.

\bibitem{mt}
OPAL Collaboration, K.~Ackerstaff {\em et al.}, Eur.~Phys.~J.~{\bf
C2} (1998) 213.

\bibitem{pythia99}
T.~Sj\"ostrand, update release notes
\\ (Internet communication of May 22, 2000:
{\tt http://www.thep.lu.se/\~{}torbjorn/Pythia.html}).

\bibitem{gopal}
J.~Allison {\em et al.}, Nucl.~Instr.~and Meth.~{\bf A317} (1992) 47.

\bibitem{lepew98}
The LEP Electroweak Working group, CERN-EP/99-015.

\bibitem{gtocc}
OPAL Collaboration, G.~Abbiendi {\em et al.}, Eur.~Phys.~J.~{\bf C13} 
(2000) 1.

\bibitem{delphicol}
DELPHI Collaboration, P.~Abreu {\em et al.}, Z.~Phys.~{\bf C59} (1993) 357.

\bibitem{durham}
S.~Catani {\em et al.}, Phys.~Lett.~{\bf B269} (1991) 432;
\\ N.~Brown and W.~J.~Stirling, Z.~Phys.~{\bf C53} (1992) 629.

\bibitem{opalrb97}
OPAL Collaboration, K.~Ackerstaff {\em et al.}, Z.Phys.~{\bf C74}
(1997) 1.

\bibitem{opalrb98}
OPAL Collaboration, G.~Abbiendi {\em et al.}, Eur.~Phys.~J.~{\bf C8}
(1999) 217.

\bibitem{herwig}
G.~Marchesini, B.~R.~Webber, G.~Abbiendi, I.~G.~Knowles,
M.~H.~Seymour, and L.~Stanco, Comp.~Phys.~Comm.~{\bf 67} (1992) 465;
\\ M.~H.~Seymour, private communication.

\bibitem{ariadne}
L.~L\"onnblad, Comp.~Phys.~Comm.~{\bf 71} (1992) 15.

\bibitem{ballestrero99} A.~Ballestrero, private communication.

\bibitem{pdg}
Particle Data Group, C.~Caso {\em et al.}, Eur.~Phys.~J.~{\bf C3}
(1998) 1.

\bibitem{sjostrandmb} T.~Sj\"ostrand, private communication.

\bibitem{jade}
JADE collaboration, W.~Bartel {\em et al.}, Z.~Phys.~{\bf C33} (1986) 23;
\\ JADE collaboration, S.~Bethke {\em et al.}, Phys.~Lett.~{\bf B213}
(1988) 235.

\bibitem{lephfwg}
The LEP collaborations, ALEPH, DELPHI, L3 and OPAL,
Nucl.~Instr.~Meth.~{\bf A378} (1996) 101;
\\ Updated averages as described in `Presentation of LEP Electroweak
Heavy Flavour Results for Summer 1998 Conferences', LEPHF 98-01
\\ (Internet communication of May 22, 2000:
{\tt http://www.cern.ch/LEPEWWG/heavy/}).

\bibitem{peterson}
C.~Peterson, D.~Schlatter, I.~Schmitt and P.~Zerwas, Phys.~Rev.~{\bf
D27} (1983) 105.

\bibitem{collins}
P.~Collins and T.~Spiller, J.~Phys.~{\bf G11} (1985) 1289.

\bibitem{kartvel}
V.~G.~Kartvelishvili, A.~K.~Likhoded and V.~A.~Petrov,
Phys.~Lett.~{\bf B78} (1978) 615.

\bibitem{markiii}
MARK III collaboration, D.~Coffman {\em et al.}, Phys.~Lett.~{\bf
B263} (1991) 135.

\end{thebibliography}
\end{document}